\documentclass[sigconf]{acmart}
\AtBeginDocument{%
  
}

\copyrightyear{2026}
\acmYear{2026}
\setcopyright{cc}
\setcctype{by}
\acmConference[CCS '26]{Proceedings of the 2026 ACM SIGSAC Conference on Computer and Communications Security}{November 15--19, 2026}{The Hague, Netherlands}
\acmBooktitle{Proceedings of the 2026 ACM SIGSAC Conference on Computer and Communications Security (CCS '26), November 15--19, 2026, The Hague, Netherlands}
\acmDOI{10.1145/3830454.3832585}
\acmISBN{979-8-4007-2871-6/2026/11}

\newif\ifshownewblue
\shownewbluefalse
\newcommand{\newtext}[1]{%
  \ifshownewblue
    \begingroup\color{blue}#1\endgroup
  \else
    #1%
  \fi
}

\usepackage[linesnumbered,ruled,vlined]{algorithm2e} 
\usepackage{multirow} 
\usepackage{makecell}
\usepackage{booktabs} 
\usepackage{threeparttable} 
\usepackage{xcolor, xspace}
\usepackage{pifont} 
\usepackage{subcaption}
\usepackage{xspace}
\usepackage[dvipsnames,table,xcdraw]{xcolor}
\def\eg{\emph{e.g.,}\xspace}
\def\etc{\emph{etc.}\xspace}
\def\ie{\emph{i.e.,}\xspace}
\def\basename{TurboRetry}
\def\sysname{\basename\xspace}
\def\slowsysname{\basename$^{\dagger}$\xspace}

\def\sect{\S}
\def\fig{Figure}
\def\yes{\textcolor{NavyBlue}{\ding{51}}}
\def\no{\textcolor{RedOrange}{\ding{55}}}
\def\retry{\texttt{Retry}\xspace}

\newcommand{\one}{({i})\xspace}
\newcommand{\two}{({ii})\xspace}

\def\eg{e.g.,\xspace}
\def\etc{\emph{etc.}\xspace}
\def\ie{\emph{i.e.,}\xspace}
\def\sect{Section}

\begin{document}

\title{\sysname: Mitigating Large-Scale QUIC Handshake Floods with Off-the-Shelf DPU Offloading}


\author{Jiahao Wu}
\orcid{0009-0007-7248-3071}
\affiliation{%
  \institution{Institute of Computing Technology,
Chinese Academy of Sciences}
  \city{Beijing}
  \country{China}
}
\affiliation{%
  \institution{University of Chinese Academy of Sciences}
  \city{Beijing}
  \country{China}
}
\email{wujiahao15@mails.ucas.ac.cn}

\author{Heng Pan}
\authornote{Corresponding author.}
\orcid{0000-0002-5506-5958}
\affiliation{%
  \institution{Computer Network Information Center, Chinese Academy of Sciences}
  \city{Beijing}
  \country{China}
}
\email{panheng@cnic.cn}

\author{Kai Lv}
\orcid{0009-0005-7675-5711}
\affiliation{%
  \institution{Institute of Computing Technology,
Chinese Academy of Sciences}
  \city{Beijing}
  \country{China}
}
\affiliation{%
  \institution{University of Chinese Academy of Sciences}
  \city{Beijing}
  \country{China}
}
\email{lvkai20z@ict.ac.cn}

\author{Zhenyu Li}
\orcid{0000-0002-9959-1124}
\affiliation{%
  \institution{Institute of Computing Technology,
Chinese Academy of Sciences}
  \city{Beijing}
  \country{China}
}
\affiliation{%
  \institution{University of Chinese Academy of Sciences}
  \city{Beijing}
  \country{China}
}
\email{zyli@ict.ac.cn}

\author{Yanbiao Li}
\orcid{0000-0001-7408-2203}
\affiliation{%
  \institution{Computer Network Information Center, Chinese Academy of Sciences}
  \city{Beijing}
  \country{China}
}
\affiliation{%
  \institution{University of Chinese Academy of Sciences}
  \city{Beijing}
  \country{China}
}
\email{lybmath@cnic.cn}

\author{Gaogang Xie}
\orcid{0000-0003-4964-1135}
\affiliation{%
  \institution{Computer Network Information Center, Chinese Academy of Sciences}
  \city{Beijing}
  \country{China}
}
\affiliation{%
  \institution{University of Chinese Academy of Sciences}
  \city{Beijing}
  \country{China}
}
\email{xie@cnic.cn}


\renewcommand{\shortauthors}{Wu et al.}

\begin{abstract}


The modern transport protocol QUIC is designed to enhance network performance and security, but it remains vulnerable to handshake flooding attacks. Such attacks exhaust CPU resources by forcing the server to perform expensive cryptographic operations via a large number of handshaking requests. QUIC provides a built-in defense mechanism, the Retry mechanism, to mitigate these attacks. However, our experiments reveal that it can still become a performance bottleneck under large-scale QUIC handshake floods due to substantial computational overhead.

In this paper, we design and implement TurboRetry, a split design, that offloads the Retry mechanism onto DPUs to efficiently mitigate QUIC handshake floods. TurboRetry partitions the tasks of the Retry into two categories, and then assigns them to the DPUs and the host, respectively. To preserve QUIC semantics and reduce the coordination overhead, TurboRetry designs an extended Retry token format and an efficient cooperation scheme. In addition, TurboRetry offloads the connection authorization task to the on-path DPA to further improve both performance and security. Our evaluation shows that TurboRetry outperforms the host-side implementation by a wide margin, improving throughput by 10–20$\times$.
\end{abstract}





\maketitle

\section{Introduction}

QUIC (Quick UDP Internet Connections) is a modern transport protocol built on UDP that combines the features of HTTP/2 and TLS, serving as the foundation of HTTP/3~\cite{gquic, rfc9000, http3rfc, quic-security}. Its widespread adoption has made it a target for Distributed Denial-of-Service (DDoS) attacks against HTTP/3 servers~\cite{cve-2024-53259, cve-2025-29785, cve-2025-29908}. In particular, the connection establishment process has emerged as a critical attack surface, which is called QUIC handshake floods~\cite{quicsand, revisiting-quic-attacks, quic-security, quic-security-survey, quicforge, quic-attack-survey, quic-analysis}. In such attacks, adversaries send massive volumes of initial handshake packets to overwhelm servers, causing CPU-intensive cryptographic operations during the TLS-based handshake to become a major performance bottleneck~\cite{rfc9001, quic-nic-offload}.

QUIC introduces the \texttt{Retry}\footnote{To avoid confusion, we denote the retry mechanism as \retry and the corresponding packet as the Retry packet.} mechanism as a built-in defense against handshake floods~\cite{rfc9000}. Its core idea is to validate the source address of a client before performing the computationally expensive cryptographic handshake. Although the \texttt{Retry} mechanism indeed improves resilience against handshake floods~\cite{quicsand}, it can still become a bottleneck under large-scale QUIC handshake floods (see \sect~\ref{subsec:existing-retry-lim}). This is because it 
requires several operations, including token generation, integrity tag computation, and token verification, which impose additional computational overhead that does not scale well under heavy traffic.
Consequently, offloading \retry to hardware emerges as a promising approach, with several drafts being proposed~\cite{ietf-quic-retry-offload}. However, existing proposals either incur significant overhead due to encryption key synchronization or fail to satisfy required security guarantees.

Recently, Data Processing Units (DPUs), such as BlueField-3~\cite{bf3}, provides a promising platform for offloading \texttt{Retry}. By extending network interface cards with integrated CPU cores and hardware look-aside accelerators~\cite{bf3-acc-compress,bf3-sketch},
DPUs offer high programmability, high-bandwidth interfaces
and efficient processing of encrypted traffic
(\eg Pigasus~\cite{pigasus}).
Offloading defense mechanisms to DPUs enables the filtering of large volumes of attack traffic before it reaches the host, thereby effectively protecting critical services.

Despite this potential, designing a practical offloading system to efficiently defend against QUIC handshake flood attacks is challenging, as a DPU-based Retry mechanism must carefully balance task allocation, maintain protocol correctness, and ensure high performance.
First, careful task partitioning is required because some Retry operations are poorly suited for the DPU’s relatively weaker processors, while others can benefit from hardware look-aside accelerators. The partitioning must consider both security and performance constraints to maximize throughput.
Second, the correctness and consistency of the protocol must be preserved, as the Retry mechanism is tightly coupled with the QUIC protocol stack, and all connection states must be synchronized between the DPU and the host stack to ensure correct connection establishment.
Third, most DPU processors and hardware accelerators reside off the fast path, requiring packets to be forwarded from the on-path pipeline to off-path units, which introduces significant additional latency. Minimizing this latency while maintaining high throughput is therefore a challenge.

To cope with these challenges, we propose a defense system \sysname that adopts a split design to offload the QUIC \retry mechanism onto DPUs, thereby achieving high-performance mitigation against QUIC handshake floods.
To this end, we introduce a universal token format, secured with AES-GCM encryption, to enhance protection. We partition the \retry mechanism into two categories of tasks: \one \emph{stateless tasks} (\ie token generation, integrity tag computation, and token verification), and \two \emph{stateful tasks} (\ie connection management). Stateless tasks are well-suited for acceleration by DPU hardware accelerators, whereas stateful tasks are processed by the more powerful host processors.
To preserve protocol correctness, we design an efficient cooperation scheme between the host and DPUs, leveraging the universal token to synchronize connection states while maintaining compliance with QUIC semantics.
We further offload the connection authorization task to the data path accelerator (DPA), an on-path component in DPUs, which directly forwards verified flows to the host with low transmission latency while proactively discarding unverified flows to enhance protection.
We also support the 0-RTT feature of QUIC via the universal token design.
Experimental results demonstrate that \sysname achieves strong resilience against adaptive adversaries, consistently maintaining reliable delivery of legitimate traffic under large-scale handshake flood attacks.
Compared to a purely host-side implementation, \sysname delivers a 10--20$\times$ throughput improvement while introducing only a negligible delay, and gracefully falls back to the host-based solution upon failures without disrupting service availability.

In summary, our primary contributions are as follows:
\begin{itemize}
    \item We propose a defense system, \sysname, with a split design (\sect~\ref{subsec:sys-arch}) that offloads the QUIC \retry mechanism onto DPUs (\sect~\ref{sec:token-design}) to efficiently mitigate large-scale QUIC handshake floods.
    \item We devise an efficient cooperation scheme (\sect~\ref{sec:host-cooperation}) where DPUs and the host coordinate via encrypted tokens to synchronize connection states without violating QUIC semantics. In addition, we offload the connection authorization task to the DPA (\sect~\ref{subsec:dpa-cache}), which directly forwards verified flows to the host while proactively discarding unverified flows, thereby improving both performance and security.
    \item We implement the \sysname prototype on the BlueField 3 DPU and make it publicly available~\cite{turboretry-repo}. Experiments show that it is resilient against attack rates of 3 Mpps \textit{without any packet loss}, gains 10--20$\times$ throughput improvement over software solutions, and only introduces minimal additional latency (\sect~\ref{sec:evaluation}).
\end{itemize}

We begin by outlining the limitations of the existing \retry mechanism in \sect~\ref{sec:motivation}, and then present the threat model, key challenges, and system architecture of \sysname in \sect~\ref{sec:overview}. A security analysis is provided in \sect~\ref{sec:security-analysis}. We further discuss additional features and potential future directions in \sect~\ref{sec:discussion}, present related work in \sect~\ref{sec:related-work}, and conclude in \sect~\ref{sec:conclusion}.

\section{Background and Motivation}\label{sec:motivation}
In this section, we first recall the key features of QUIC, with a focus on its handshake and retry mechanisms. We then further analyze the limitations of existing retry mechanisms and discuss the new opportunities enabled by DPUs.

\subsection{QUIC Handshake and Retry Mechanism}\label{subsec:handshake}
QUIC is designed to improve the performance and security of network transmission~\cite{quic-security}.
It merges TCP's three-way handshake and TLS negotiation into a single handshake~\cite{rfc9000, rfc9001}, significantly reducing connection establishment latency.
QUIC’s mandatory end-to-end encryption hides protocol metadata from on-path devices, effectively mitigating the security risks associated with plaintext transmission in TCP~\cite{tcp-middlebox-security}.


\textbf{QUIC packet formats.}
QUIC packets are carried inside UDP payloads and come in two forms: long-header and short-header packets. Long-header packets are used for connection establishment, involving Initial, Retry, Handshake, 0-RTT, and Version Negotiation packets.
Short-header packets (\ie 1-RTT packets) are used for normal data transmission.
Note that the payloads of Initial, Handshake and 1-RTT packets are encrypted using Authenticated Encryption with Associated Data (AEAD) algorithms~\cite{aead} (\eg AES-GCM~\cite{aes-gcm}).


\begin{figure*}[t]
    \centering
    \begin{subfigure}[b]{0.34\textwidth}
        \centering
        \includegraphics[width=\linewidth]{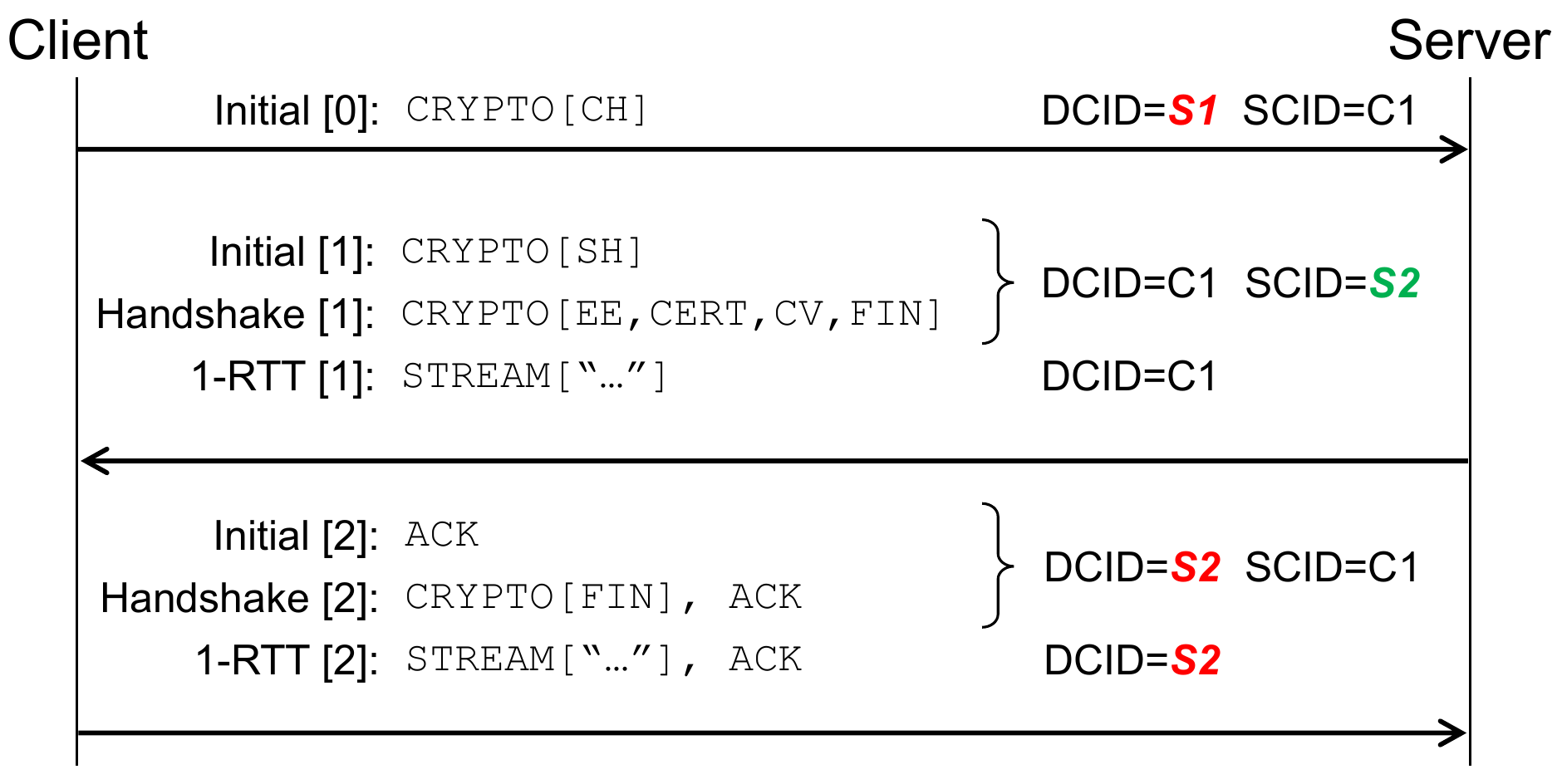}
        \caption{\newtext{1-RTT handshake.}}
        \label{fig:handshake-woretry}
    \end{subfigure}
    \hfill
    \begin{subfigure}[b]{0.33\textwidth}
        \centering
        \includegraphics[width=\linewidth]{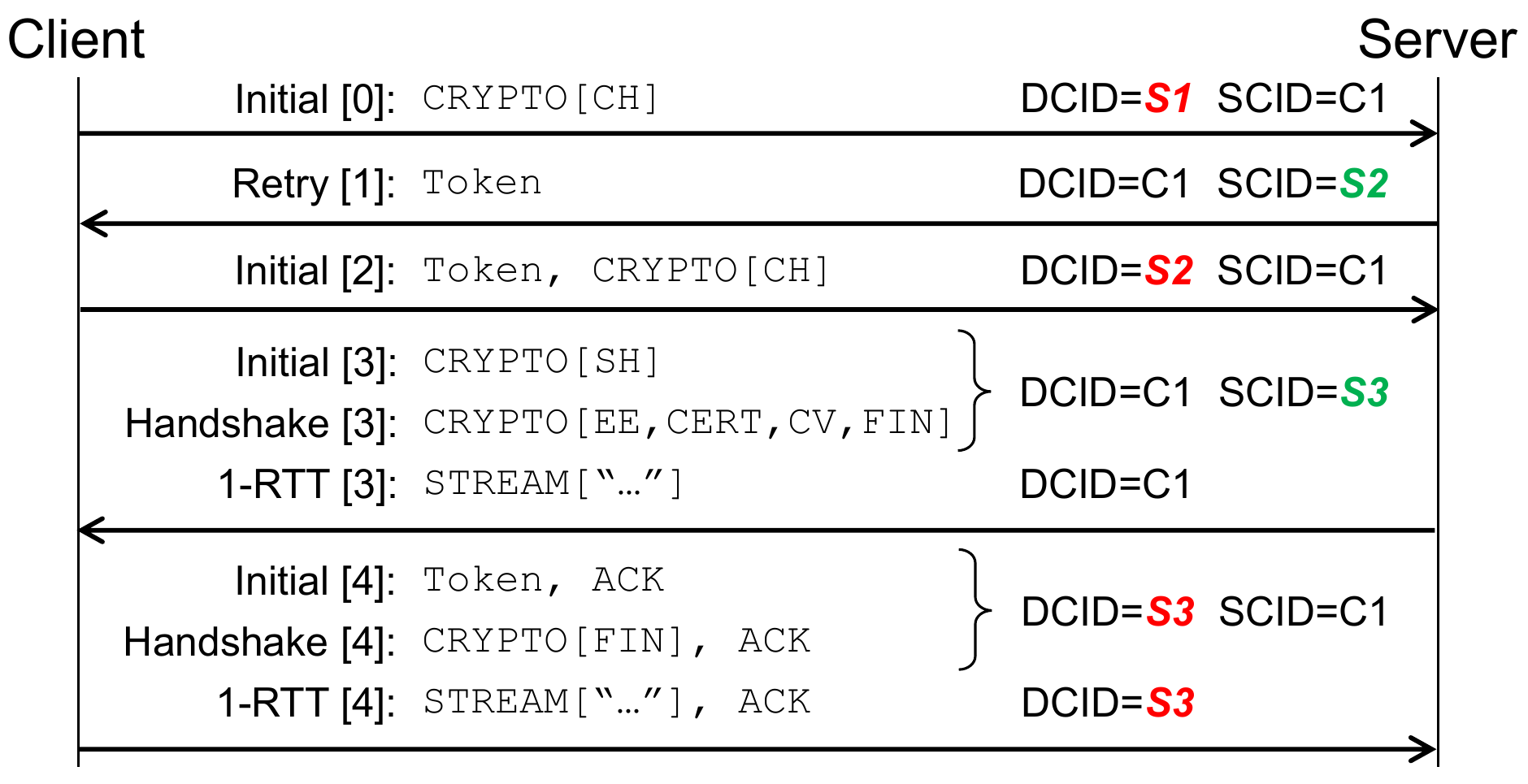}
        \caption{\newtext{1-RTT handshake with \retry.}}
        \label{fig:handshake-wretry}
    \end{subfigure}
    \hfill
    \begin{subfigure}[b]{0.3\textwidth}
        \centering
        \includegraphics[width=\linewidth]{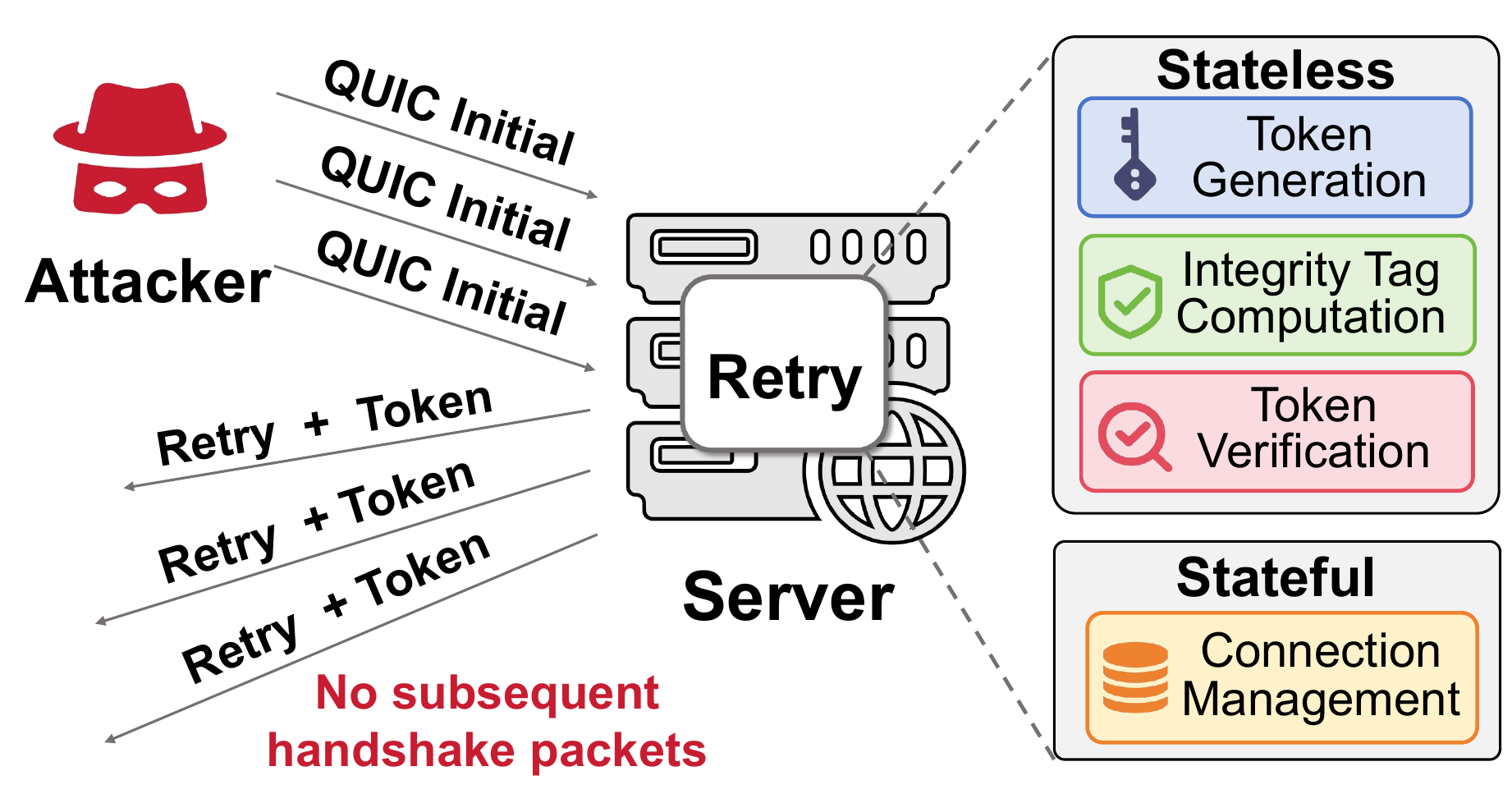}
        \caption{Main functions of \retry.}
        \label{fig:host-retry}
    \end{subfigure}
    \caption{QUIC handshake and the \retry mechanism. \newtext{The numbered labels indicate the order of packet-exchange rounds.}}
    \label{fig:1rtt-handshake}
\end{figure*}

\textbf{QUIC handshake mechanism.}
A typical QUIC handshake comprises three main steps (\fig~\ref{fig:handshake-woretry}).
\textbf{\ding{172}} The client sends an Initial packet containing a TLS ClientHello (CH) message encrypted in its payload.
\textbf{\ding{173}} Upon receiving this packet, the server decrypts its payload and replies with an Initial packet containing a TLS ServerHello (SH) message.
The server then sends a Handshake packet with its TLS handshake messages—Encrypted Extensions (EE), Certificate (CERT), and Certificate Verify (CV)—followed by a 1-RTT packet containing application data.
\textbf{\ding{174}} The connection is established once the server receives a sequence of packets from the client, including an Initial packet with ACK, a Handshake packet with the client's TLS information, and a 1-RTT packet.

\textbf{Retry mechanism.}
\retry~\cite{rfc9000} is a built-in defense designed to mitigate QUIC handshake floods and has been widely adopted by QUIC implementations listed by the QUIC Working Group~\cite{quic-impl-list}.
As illustrated in \fig~\ref{fig:handshake-wretry}, \retry introduces two additional steps prior to connection establishment:
\textbf{\ding{172}} The server responds to the client's first Initial packet with a Retry packet containing a stateless token (\texttt{Token}), without decrypting the payload;
\textbf{\ding{173}} The client retransmits the Initial packet carrying the token (\texttt{Token}) \newtext{in the header}.
The server proceeds with the handshake only after successfully verifying the token. 
By deferring decryption and other expensive cryptographic operations, \retry prevents CPU exhaustion caused by spoofed Initial packets.
Note that clients are required to include the retry token $T$ in all the subsequent Initial packets to prove its authenticity.
Although \retry introduces additional interaction delay, it provides strong protection against QUIC handshake floods~\cite{quicsand}.

We summarize the main operations involved in \retry into two groups: stateless and stateful functions (shown in \fig~\ref{fig:host-retry}).
Stateless functions include token generation, Retry packet integrity tag computation, and token verification, which respectively generate tokens for client validation, protect Retry packets against tampering using AEAD algorithms (\eg AES-GCM), and validate returned tokens without maintaining per-connection state.
The stateful function primarily involves connection management, which requires maintaining server-side state across connections.
Among these, only token-related functions (token generation and verification) vary significantly across different implementations, as retry tokens are self-produced and self-consumed by the same server and their formats are independently defined by each implementation~\cite{rfc9000}.

\newtext{
\textbf{Connection identifier negotiation.}
QUIC decouples a connection from the traditional five-tuple by using connection identifiers (CIDs). In a long-header packet, the destination CID (DCID) identifies the intended receiver and is used for routing, while the source CID (SCID) advertises the CID that the peer should use as the DCID in subsequent packets.
Therefore, the DCID used by an endpoint is negotiated during the handshake.
Once the negotiation completes, 1-RTT short-header packets only need to carry the peer-selected CID as the DCID for routing.
To authenticate this negotiation, QUIC encodes the relevant DCID choices into the transport parameters extension, which are exchanged and authenticated as part of the TLS handshake to prevent CID tampering or \retry forgery.

In the 1-RTT handshake (\fig~\ref{fig:handshake-woretry}), the client's DCID
switches from its initial random value \texttt{S1} to the server-selected CID \texttt{S2}, which is advertised as the server's SCID during the handshake. Accordingly, \texttt{S1} and \texttt{S2} are encoded in the server's transport parameters.
When \retry is used (\fig~\ref{fig:handshake-wretry}), the DCID transition includes an additional Retry-issued CID. The client first uses \texttt{S1}, the original DCID (ODCID), then switches to the Retry SCID (RSCID) \texttt{S2} when retransmitting the Initial packet with the retry token, and finally uses the server's post-validation SCID \texttt{S3} for subsequent packets. In this case, \texttt{S1}, \texttt{S2}, and \texttt{S3} are encoded in the server's transport parameters.
}


\textbf{Version negotiation and Zero Round-Trip Time (0-RTT) resumption.}
Version negotiation is triggered when a client proposes an unsupported QUIC version, causing the server to negotiate a mutually supported version before the standard 1-RTT handshake.
0-RTT resumption reduces connection setup latency by allowing a client to send encrypted application data in its first flight using cryptographic parameters from a prior connection.
It is commonly combined with a server-issued \texttt{NEW\_TOKEN}, which lets the client prove prior address validation and avoid being downgraded by an additional Retry exchange.
\newtext{However, because 0-RTT sends data before full authentication and key confirmation, it also expands the attack surface~\cite{rfc8446}.}


\subsection{Limitations of the Existing Retry}\label{subsec:existing-retry-lim}
We next introduce the limitations of existing \retry mechanisms, including host-based \retry mechanisms and \retry offloading drafts.

\textbf{Host-based \retry solutions.}
Host-based \retry solutions refer to QUIC implementations built on the host UDP stack that support the \retry mechanism.
As mentioned before, one of their major differences lies in the token format, which provides varying levels of security.
More complex token formats incur higher computational overhead, resulting in reduced \retry throughput.
We evaluate the maximum attack rate that the host-based \retry can withstand using different token formats selected from~\cite{quic-impl-list}.
In our experiment, we use Quiche~\cite{quiche}, a Rust-based QUIC implementation, as the test platform. 
Specifically, we customize Quiche to implement five distinct token formats: plaintext, HMAC-authenticated, AES-GCM-encrypted, RSA-encrypted, and Ed25519-signed tokens.
The plaintext format directly embeds the source IP address, source port, and source connection identifier into the token without cryptographic protection. The HMAC-based format computes the hash value of the plaintext content and appends the resulting hash to the end.
The AES-GCM-based format encrypts the plaintext token body.
The Ed25519-based format signs the token body with the server's private key and appends the signature; validation verifies the signature using the corresponding public key.
The RSA-based format encrypts the same token body using an RSA key pair.
We adopt the experimental setup described in \sect~\ref{subsec:expr-setup} and use the successful request ratio as the metric to characterize the server's state under attack.

\begin{figure}[hbp]
    \centering
    \includegraphics[width=0.78\linewidth]{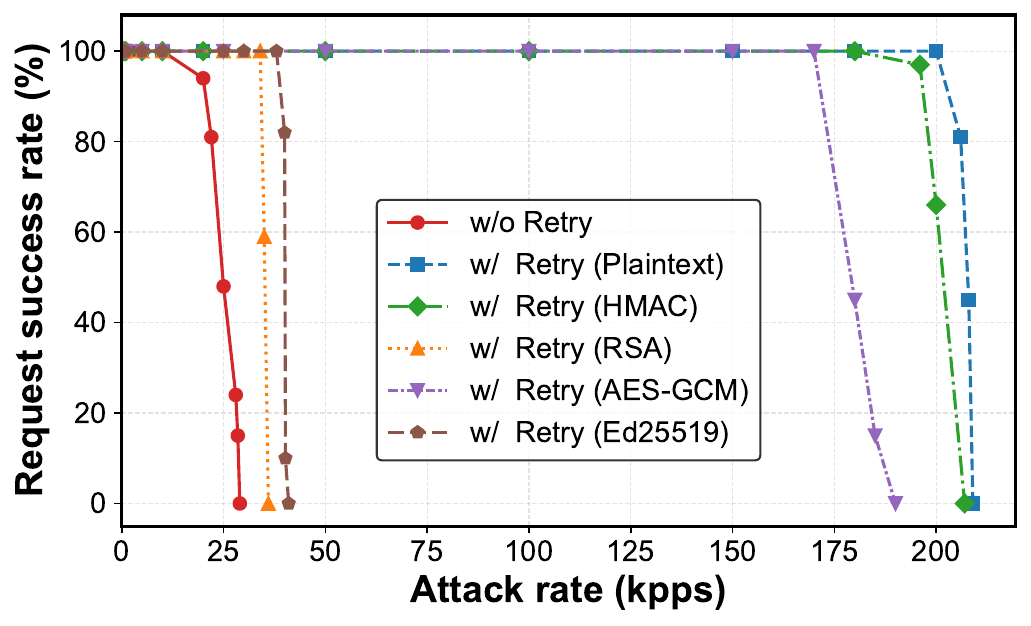}
    \caption{\newtext{\retry performance measurement. Note that the attack rate is measured in kilo packets per second (kpps).}}
    \label{fig:aioquic-brr}
\end{figure}

\textbf{Limitations of host-based \retry.}
As depicted in \fig~\ref{fig:aioquic-brr}, the basic QUIC handshake without \retry can withstand attacks of up to 28 kpps. When \retry is enabled, plaintext tokens, HMAC-based tokens, AES-encrypted tokens, Ed25519-signed tokens, and RSA-encrypted tokens can mitigate attack rates of up to 209 kpps, 207 kpps, 188 kpps, 38 kpps, and 32 kpps, respectively. These results demonstrate that enabling \retry improves DDoS resistance by 1.1--7.5$\times$. 
However, the host-based \retry can still become a performance bottleneck under large-scale attacks ($>209$ kpps), and increasing the computational complexity of token-protection operations can significantly reduce \retry throughput (\ie defense efficiency). Therefore, the community has turned to offloading \retry to hardware to balance token security and \retry performance.


\begin{table}[htbp]
    \centering
    \caption{Comparison between the \retry offloading draft and \sysname.}
    \label{tbl:cmp-prior}
    \begin{threeparttable}
    \small 
    \begin{tabular}{c|c|c|c}
    \toprule
    Solution  & Fail-open & Easy-to-deploy & 0-RTT \\
    \midrule
    No-shared-state~\cite{ietf-quic-retry-offload} & \no & \yes & \no \\
    Shared-state~\cite{ietf-quic-retry-offload} & \yes & \no & \yes \\
    \sysname  & \yes & \yes & \yes \\
    \bottomrule
    \end{tabular}
    \end{threeparttable}
\end{table}
Overall, a practical \retry offloading solution should satisfy three key requirements: (1) it should operate in a \textbf{fail-open} manner to preserve service availability when the offloading device fails or becomes unavailable, (2) it should be \textbf{easy-to-deploy} without complex host--device coordination, and (3) it should support the prominent \textbf{0-RTT} resumption feature, rather than forcing resumed connections to fall back to a full 1-RTT handshake.
Table~\ref{tbl:cmp-prior} summarizes two approaches proposed in the existing \retry offloading draft~\cite{ietf-quic-retry-offload}: no-shared-state and shared-state offloading.

\newtext{
In \textbf{no-shared-state} offloading, the device independently generates and validates plaintext Retry tokens without sharing state with the host.
This design keeps the data path simple and avoids host--device synchronization.
However, since the device does not share the server's token format or cryptographic state, it cannot recognize server-issued \texttt{NEW\_TOKEN}s.
As a result, 0-RTT packets carrying such tokens are dropped at the offloading device, forcing QUIC's low-latency 0-RTT resumption to degrade into a full 1-RTT handshake.
The design also makes availability depend on the offloading device. When the device fails, the host cannot directly take over token validation, causing the system to fail closed.

In \textbf{shared-state} offloading, the device and the host share the cryptographic state needed to generate and validate encrypted retry tokens.
This enables both sides to validate each other's tokens and allows the device to support 0-RTT resumption with \texttt{NEW\_TOKEN}.
However, this benefit comes at the cost of deployability.
The system must synchronize keys, nonces, and rotation epochs between the host and the device.
More importantly, to avoid security vulnerabilities caused by nonce reuse in AES-GCM, the system ought to adopt an aggressive key-rotation strategy, such as enforcing key rotation after every $2^{23}$ packets.
Such frequent multi-state management and synchronization introduce substantial operational overhead, creating significant engineering barriers to deployment and scaling in practical network environments.
Thus, shared-state offloading turns \retry offloading into a distributed cryptographic state-management problem.

In summary, existing designs expose a tradeoff: no-shared-state offloading is easy to deploy but fails closed and breaks 0-RTT, whereas shared-state offloading supports 0-RTT but requires complex and frequent key and nonce synchronization.
\sysname bridges this gap by deriving per-connection keys and nonces from a single pre-shared master key using HKDF (see Algorithm~\ref{algo:key-gen}), eliminating complex cryptographic state distribution between the host and the offloading device.
It further introduces a universal retry token format (mentioned in Section~\ref{sec:token-design}) that supports both Retry validation and \texttt{NEW\_TOKEN}-based 0-RTT resumption, preserving simple deployment without sacrificing QUIC's low-latency resumption feature.
}
\subsection{Opportunities of Data Processing Units}\label{subsec:bf3-arch}

A Data Processing Unit (DPU), also known as a Smart Network Interface Card (SmartNIC), is a high-performance programmable system-on-chip (SoC) processor that integrates a multi-core CPU, network interfaces, and hardware accelerator engines.
It can replace traditional NICs and offer performance improvements by offloading specific tasks (\eg networking, storage, security, \etc) from host CPU cores to the DPU.
Table~\ref{tbl:comp-dpus}\footnote{Table entries marked with '-' indicate that the corresponding information is not publicly available.} compares the cost and key capabilities of representative off-the-shelf DPUs, highlighting the trade-offs between device functionality and deployment cost.
DPUs have emerged as a promising solution for data centers and cloud environments~\cite{dpu-is-good,dpu-is-good-2,dpu-is-good-3}. In this paper, we focus on the state-of-the-art (SOTA) DPU, NVIDIA BlueField-3~\cite{bf3}.

\begin{table}[ht]
    \centering
    \caption{Comparison of representative off-the-shelf DPUs.}
    \label{tbl:comp-dpus}
    \footnotesize 
    \setlength{\tabcolsep}{3pt} 
    \begin{tabular}{lccccc} 
        \toprule
        \textbf{Device} & 
        \textbf{Cap.} & 
        \makecell[c]{Cost\\(\$)} &     
        \makecell[c]{Power\\(W)} &     
        \makecell[c]{Data Plane\\Programmability} & 
        \makecell[c]{AES-GCM\\Accelerator} \\ 
        \midrule
        Nvidia BlueField-3 & 400Gbps & 3,455 & 75 & \yes & \yes \\
        AMD Pensando DSC3  & 400Gbps & -     & -  & \yes & \yes \\
        Intel IPU          & 200Gbps & 4,318 & 75 & \yes & \yes \\
        Marvell OCTEON     & 16$\times$50Gbps & - & 50 & \yes & \yes \\ 
        \bottomrule
    \end{tabular}
\end{table}

\begin{figure}[htbp]
    \centering
    \includegraphics[width=0.8\linewidth]{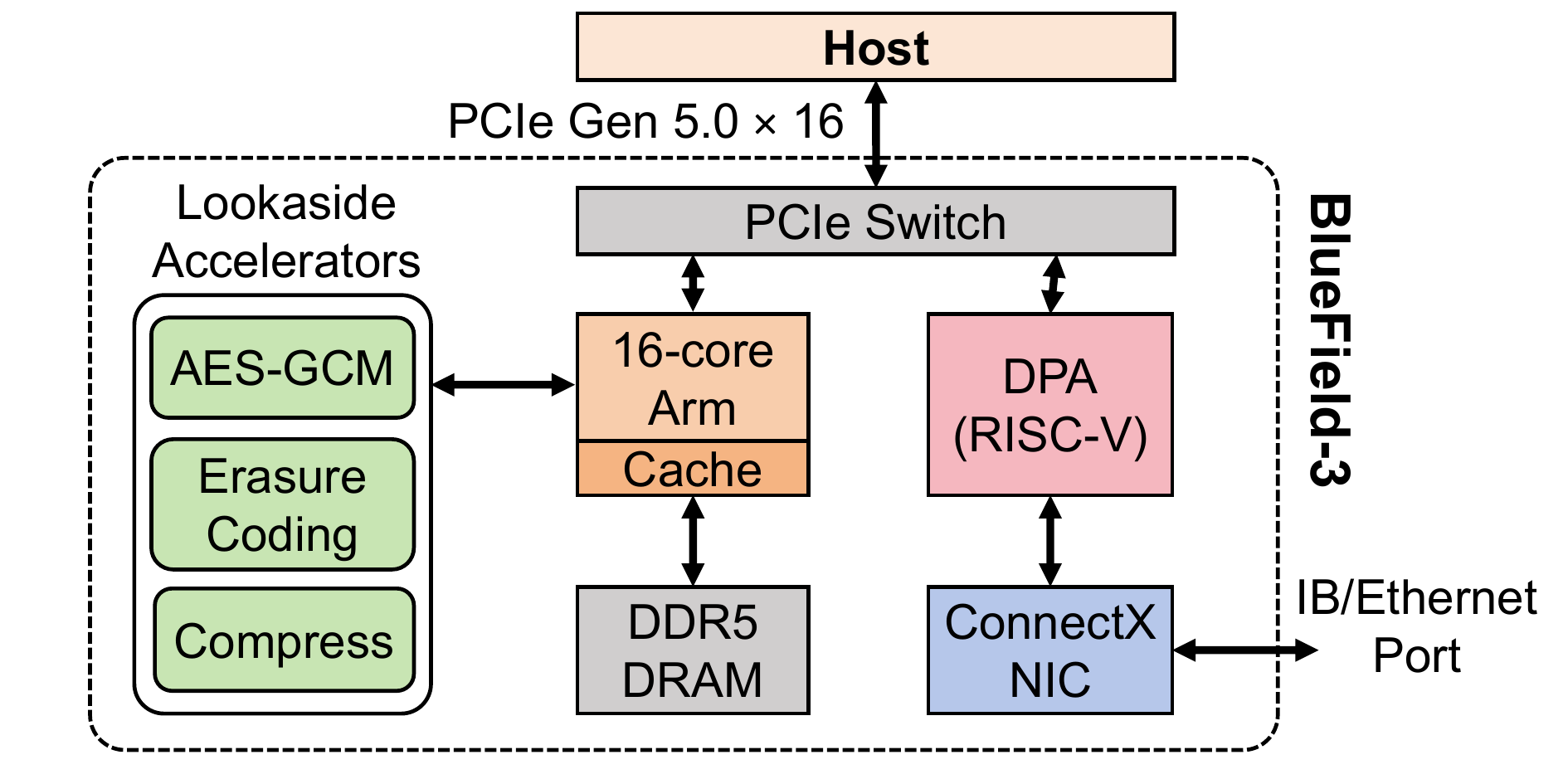}
    \caption{BlueField-3 DPU architecture.}
    \label{fig:bf3-arch}
\end{figure}

\textbf{BlueField-3 architecture.}
\fig~\ref{fig:bf3-arch} shows the overall architecture of BlueField-3 (BF3) DPU. BF3 comprises two ConnectX-7 NICs, several SoC cores and domain-specific hardware accelerators, and off-chip DDR5 memory. Each NIC supports 200/400 Gbps Ethernet or InfiniBand. The SoC includes two types of processors: (i) an off-path general-purpose processor with 16 Arm Cortex-A78 cores, and (ii) a programmable on-path Data Path Accelerator (DPA). These two processors are connected via an embedded PCIe switch. 

The DPA processor features a 16-core, 256-thread RISC-V architecture and resides on the network critical path, providing inline programmability for every incoming and outgoing packet. In addition to its own three-level cache backed by NIC private memory, the DPA can access the last-level cache (LLC) and memory of both the Arm processor and the host system via the PCIe bus using load/store instructions~\cite{dpa}.
In contrast, the Arm processor serves as a co-processor to the host CPUs, offloading computation-intensive tasks. It is connected to 32GB DDR5 memory and a suite of hardware accelerators (\eg AES-GCM engines, erasure coding units, compression modules, \etc). 

\begin{figure}[htbp]
    \centering
    \begin{subfigure}[b]{0.49\columnwidth}
        \centering
        \includegraphics[width=\linewidth]{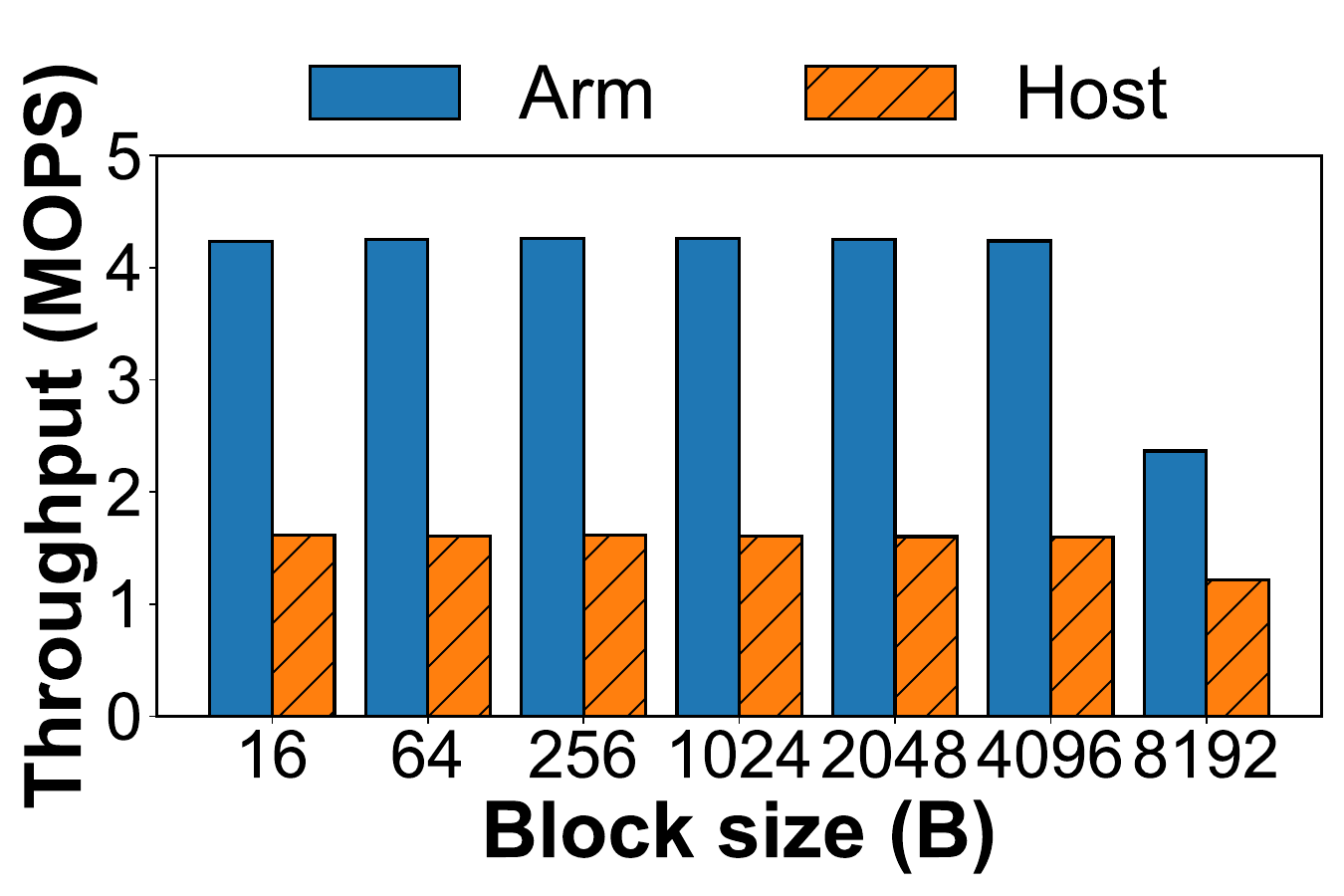}
        \caption{Throughput.}
        \label{fig:aes-throughput}
    \end{subfigure}
    \hfill
    \begin{subfigure}[b]{0.49\columnwidth}
        \centering
        \includegraphics[width=\linewidth]{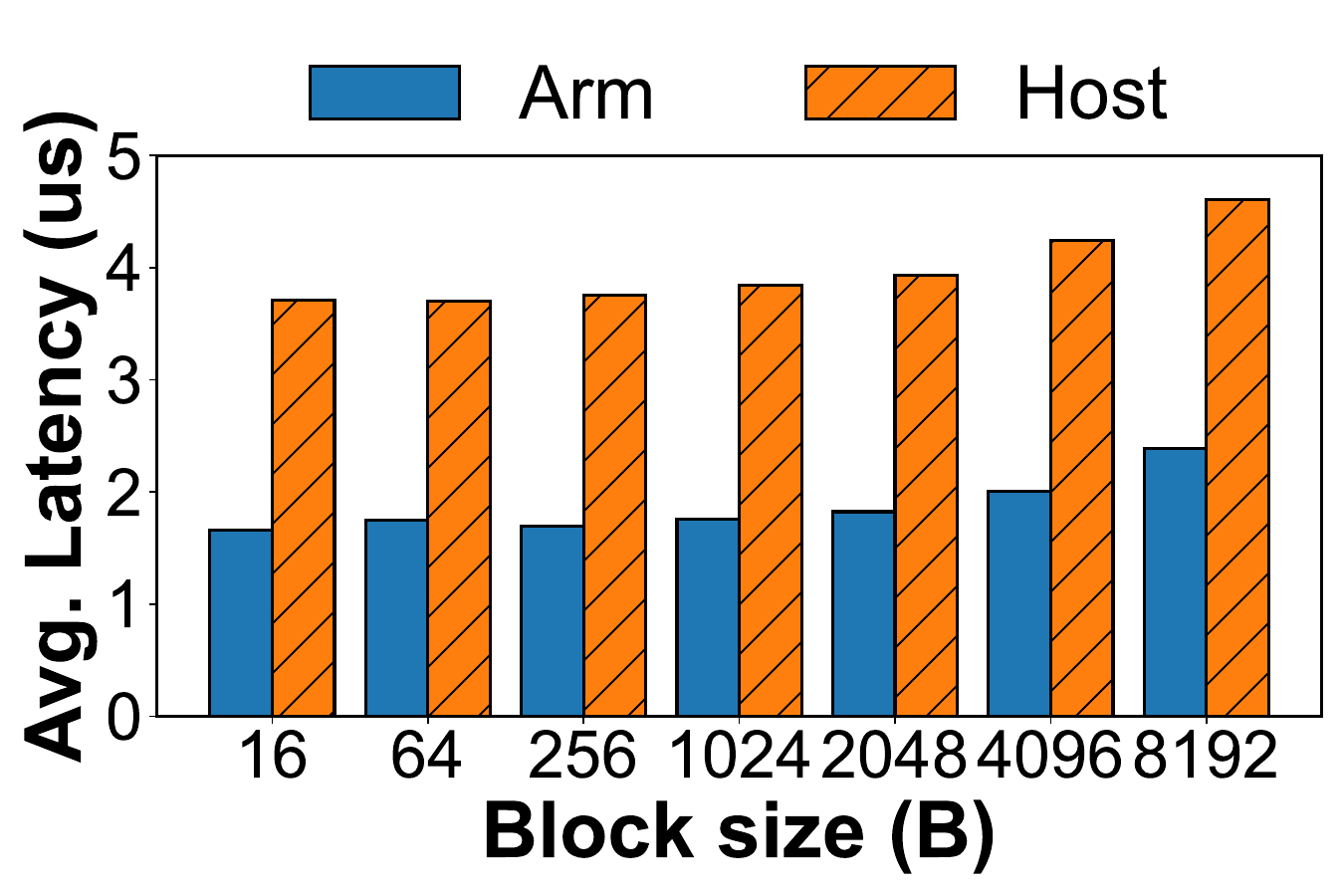}
        \caption{Latency.}
        \label{fig:aes-latency}
    \end{subfigure} 
    \caption{Benchmarking the AES-GCM hardware accelerator on BlueField-3.}
    \label{fig:aes-perf}
\end{figure}

\textbf{Benchmarking the AES-GCM accelerator.}
The QUIC handshake and data transmission involve numerous AES-GCM cryptographic operations. Although these operations can be accelerated with AES-NI~\cite{aes-ni} instructions on the host CPU, they still occupy significant CPU resources under attacks.
The AES-GCM accelerator on BF3 offers promising opportunities to both relieve CPU burden and accelerate these operations. Therefore, we measure the throughput and latency when invoking the hardware accelerator on both the host CPU and the DPU’s Arm processor using the \texttt{doca\_bench} tool. \fig~\ref{fig:aes-throughput} and \fig~\ref{fig:aes-latency} show that, compared with the host, invoking the AES-GCM accelerator on the Arm processor achieves higher throughput ($\approx$4 million operations per second for one core) and lower latency ($<2$ µs). This demonstrates the potential for offloading AES-GCM operations to BF3.

\textbf{Limitations of BlueField-3}.
Although the Arm processor provides hardware-accelerated processing capabilities that outperform host CPUs for specific tasks, it introduce additional latency because traffic must traverse the PCIe switch to reach them. 
This poses challenges in achieving both low latency and high throughput when offloading network functions from host CPUs to the DPU.

\section{\sysname Problem Setting}\label{sec:overview}

\begin{figure}[htbp]
    \centering
    \includegraphics[width=0.75\linewidth]{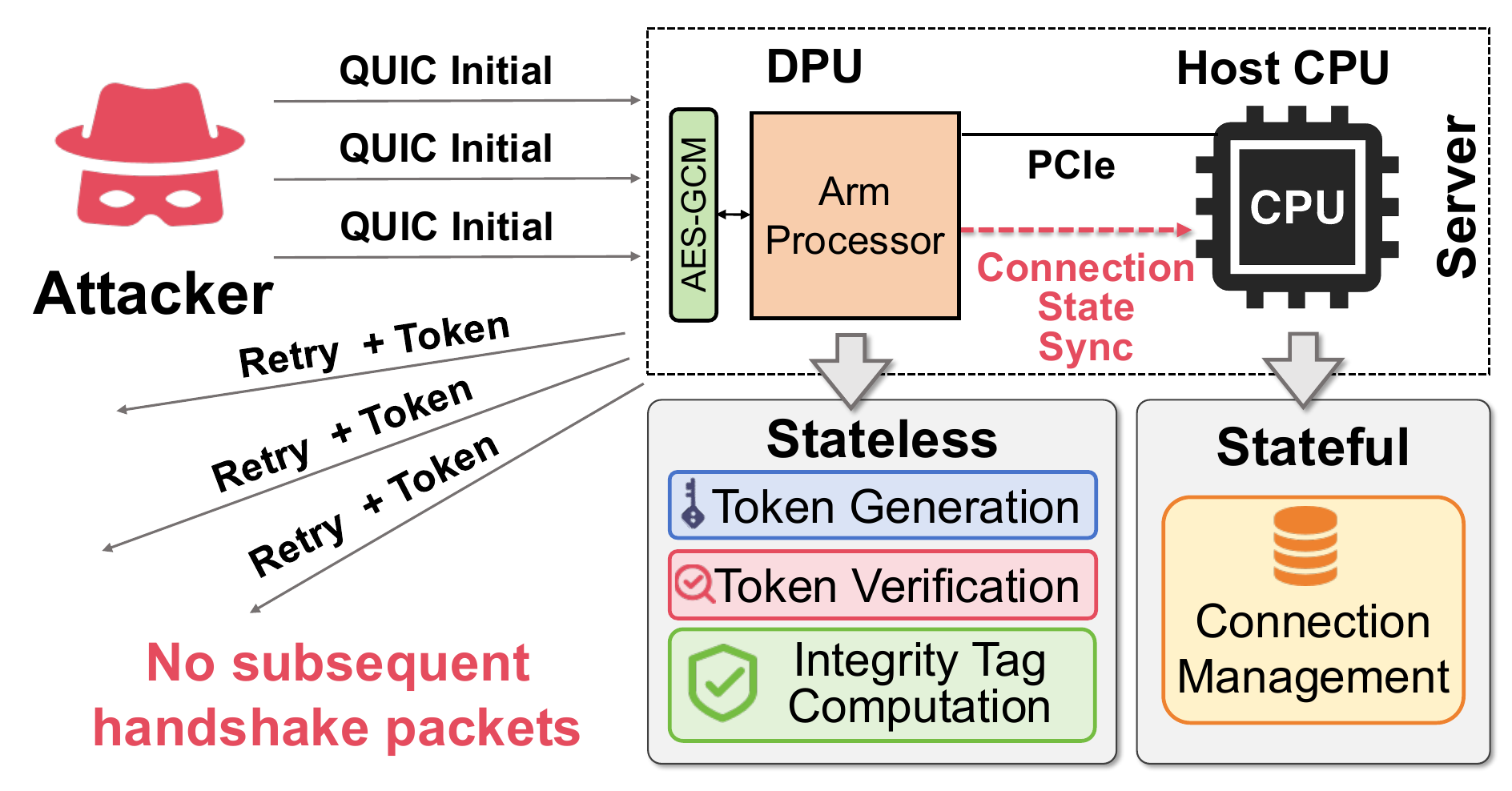}
    \caption{\newtext{Overview of DPU-based \retry.}}
    \label{fig:bf3-retry}
\end{figure}

To mitigate QUIC handshake floods, we design \sysname, a DPU-based \retry offloading mechanism (see \fig~\ref{fig:bf3-retry}). In \sysname, stateless functions are offloaded to the DPU to leverage hardware accelerators for potential performance improvement, while
stateful functions remain on the host to ensure correct maintenance of connection state within the protocol stack. Within the split design, 
the DPU synchronizes the necessary connection state with the host QUIC stack via the PCIe bus to ensure that verified connections continue to operate correctly. Under this design, we next present the threat model and the key technical challenges of \sysname.

\subsection{Threat Model}
Our threat model considers three key actors: clients, adversaries, and network providers. 

\textbf{Clients}.
We assume that benign clients follow the QUIC specification and generate well-formed connection attempts.
The primary security objective of \sysname is to protect these clients from service disruption caused by QUIC handshake flooding.
In particular, the defense must not drop or rate-limit benign traffic.

\textbf{Adversaries}.
We focus on asymmetric QUIC handshake flooding attacks, in which the adversary expends significantly fewer resources than the defender~\cite{quic-nic-offload, mcia}.
In this setting, we assume the adversary controls a botnet~\cite{marai} capable of generating massive volumes of QUIC Initial packets.
To evade simple blacklisting, the adversary employs IP spoofing, ensuring that attack traffic appears to originate from a diverse range of source IP addresses.
We assume the adversary is \textit{off-path} (or blind), meaning they cannot observe the return traffic from the victim.

\textbf{Network providers.}
We assume that the network provider manages the physical infrastructure, including servers and DPUs.
\newtext{
TurboRetry adopts a least-privilege design for provider-managed DPUs.
The DPU is delegated only the stateless address-validation logic required for QUIC Retry and is provisioned with a pre-shared token secret for generating and validating Retry tokens.
It does not possess the server's long-term TLS private keys, QUIC/TLS session keys, or application-layer secrets, and therefore cannot decrypt or forge protected QUIC application data.
We do not assume an arbitrarily malicious DPU that can exfiltrate secrets.
Instead, we consider realistic DPU-side failures, including software bugs, crashes, and transient unavailability.
The host does not blindly trust DPU forwarding decisions; instead, it revalidates the Retry token before admitting an Initial packet into the QUIC stack.
This secondary verification prevents buggy DPU behavior, stale cache entries, packet corruption, or erroneous forwarding decisions from injecting invalid connection state into the host QUIC stack.
}

\subsection{Challenges}\label{subsec:challenges}
Overall, \sysname needs to address several key technical challenges to defend against QUIC handshake flooding attacks.

\textbf{Challenge 1: Strategic partitioning of \retry under security and performance constraints.}
The \retry mechanism involves multiple tasks. Some are ill-suited for execution on the DPU due to its relatively weaker general-purpose processors compared to host CPUs, while others can benefit significantly from the DPU’s hardware accelerators.
The core challenge lies in determining an appropriate division of labor that should satisfy two constraints: (i) identifying which functions must remain on the host, and (ii) determining which computations can be safely offloaded to the DPU to achieve performance gains.
At the same time, \retry security requires encrypting its retry tokens.
However, stronger encryption improves security but inevitably increases computational overhead, degrading overall system throughput (see \sect~\ref{subsec:existing-retry-lim}).
We show how \sysname navigates these trade-offs in \sect~\ref{sec:token-design}.

\textbf{Challenge 2: Coordinating the split design without violating QUIC semantics.}
QUIC requires the host protocol stack to track all connection identifiers that the host negotiates with the client during the \retry.
The host encodes these identifiers as transport parameters in the TLS-encrypted payload of the Handshake packet and sends them to the client to complete connection setup. Omitting these parameters causes the client to reject the connection.
In our split design, these identifiers are managed on the DPU and needs to be synchronized with the host QUIC protocol stack.
Although one could introduce a new packet format to convey this information to the host, doing so would violate QUIC semantics.
\sect~\ref{sec:host-cooperation} shows how \sysname efficiently coordinates the host and the DPU to handle this challenge. 

\textbf{Challenge 3: Minimizing end-to-end transmission latency for verified connections.}
Most processors and hardware accelerators in DPUs reside off the fast path, requiring packets to be forwarded from the on-path pipeline to off-path units.
This forwarding indeed introduces significant additional transmission latency (mentioned in \sect~\ref{subsec:bf3-arch}). We show how \sysname addresses this challenge in \sect~\ref{subsec:dpa-cache}.

\section{\sysname Architecture}\label{subsec:sys-arch}

In this section, we first present an overview of \sysname, and then describe its end-to-end workflow for benign requests, including both the connection setup and the data transmission.

\subsection{Overview}

\begin{figure}[htbp]
    \centering
    \includegraphics[width=0.9\linewidth]{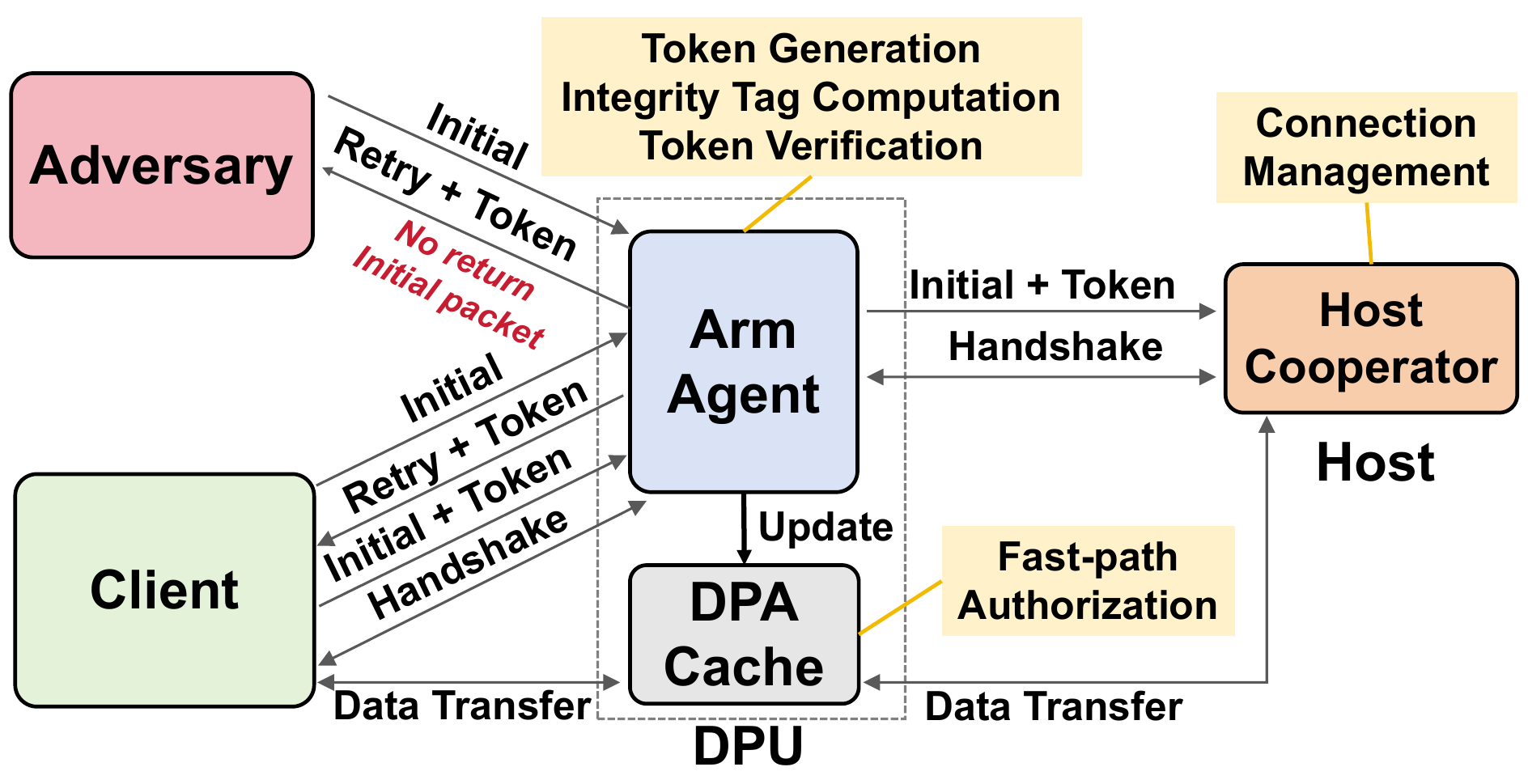}
    \caption{\newtext{\sysname architecture.}}
    \label{fig:sys-arch}
\end{figure}

As mentioned before, \sysname adopts a split design paradigm that offloads stateless functions of \retry to the DPU (including Arm and DPA processors) while managing QUIC connections through coordinated DPU–host cooperation (see \fig~\ref{fig:sys-arch}). Overall, \sysname consists of three components: Arm agent, DPA cache, and host cooperator.

\begin{figure*}[t]
    \centering
    \includegraphics[width=0.75\linewidth]{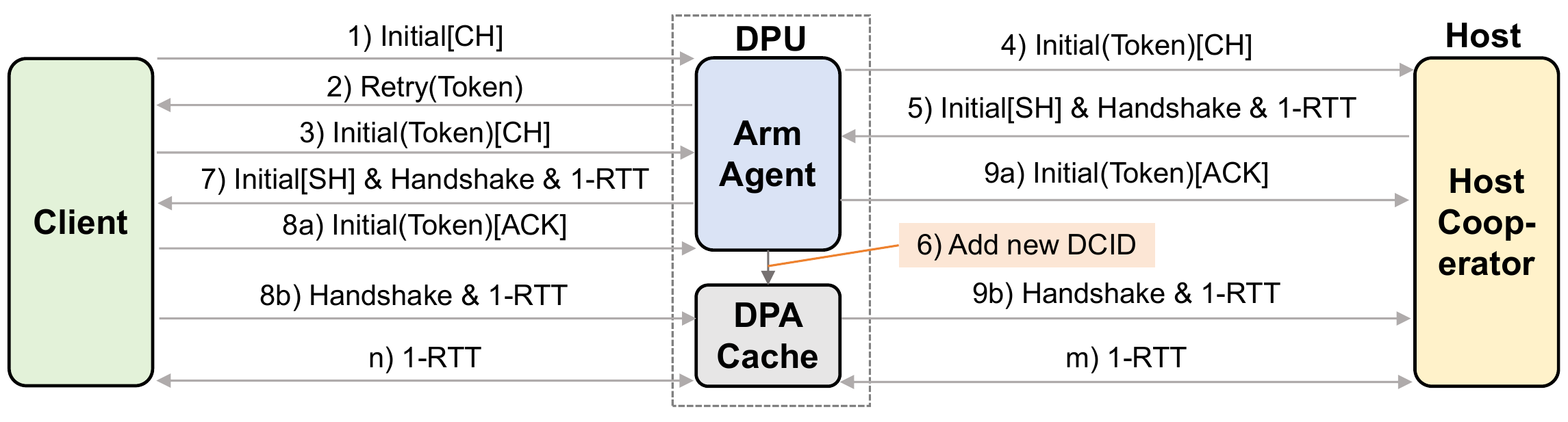}
    \caption{\newtext{End-to-end connection setup and data transfer procedure for verified connections.}}
    \label{fig:verified-setup}
\end{figure*}

\textbf{Arm agent.}
The Arm agent handles the stateless functions in \retry, including retry token generation, retry integrity tag computation, and token verification (\sect~\ref{sec:token-design}).
Note that all these functions involve AES-GCM operations. 
This design fully leverages the AES-GCM hardware accelerator directly attached to the Arm processor, significantly improving its performance.
For each Initial packet without a token, the Arm agent can generate a Retry packet with very low latency ($<2\mu$s) and withstand large-scale QUIC Handshake flooding attacks.
After successful token verification, the Arm agent forwards the client’s Initial packet to the host and installs the validated connection identifier in the DPA cache to accelerate subsequent data transmission.

\textbf{DPA cache.}
The DPA cache is primarily designed to reduce data transmission latency (\sect~\ref{subsec:dpa-cache}). 
Although the Arm agent offers greater computational capability and storage resources, its off-path architecture introduces additional latency due to PCIe switch traversal. In contrast, the on-path DPA processor resides on the network’s critical path and can steer traffic directly to the host, thereby minimizing transmission latency. Thus, the DPA cache authorizes validated connections to accelerate their data transmission and drops invalid flows before they reach the host.

\textbf{Host cooperator.}
The host-side design focuses on cooperating with the DPU to complete stateful connection management within the QUIC protocol stack for benign connections (\sect~\ref{sec:host-cooperation}). During \retry, the server negotiates its own connection identifiers with the client. To preserve the integrity of this process, QUIC requires the server to include all negotiated connection identifiers, \ie original DCID (ODCID) and retry SCID (RSCID), as transport parameters in the TLS handshake, which the client uses to verify that the server has not been hijacked. In \sysname, these identifiers are generated and maintained on the DPU and must be synchronized with the host.
To preserve QUIC semantics, we embed these two connection identifiers into the proposed secure retry token and deliver them to the host together with successfully verified Initial packets.
The host then decrypts the token, extracts the identifiers, and supplies them to the QUIC stack to continue the handshake.
Notably, both the DPU and the host 
derive cryptographic keys using the same key-derivation process, eliminating the need to explicitly share cryptographic materials for token encryption or decryption.

\subsection{End-to-end Workflow}\label{subsec:e2e-workflow}

\fig~\ref{fig:verified-setup} illustrates the complete end-to-end connection setup and data transfer procedure in \sysname for verified connections.
Upon receiving the first Initial packet from the client (\texttt{step} 1), \sysname's Arm agent generates a token according to the proposed format (\sect~\ref{sec:token-design}), computes the retry integrity tag, and delivers a Retry packet containing the token to the client (\texttt{step} 2).
In \texttt{step} 3, the Arm agent performs token verification upon receiving the client’s second Initial packet, which contains the token and ClientHello message, to authenticate the client.
If verification succeeds, the second Initial packet will be forwarded to the host immediately (\texttt{step} 4).
The host stack recovers the \texttt{RSCID} and \texttt{ODCID} from the client's Initial packet, and then generates an Initial packet containing the ServerHello message, a Handshake packet carrying the associated TLS information, and a 1-RTT packet (\texttt{step} 5).
Before these packets are transmitted to the client (\texttt{step} 7), the DPU Arm agent extracts the new connection identifier assigned by the host and updates the DPA cache (\texttt{step} 6), ensuring that subsequent packets from the client are directly steered to the host.
The Arm agent also verifies the token in the subsequent Initial packet with ACK (\texttt{step} 8a) and steers it to the host (\texttt{step} 9a).
The DPA then handles subsequent Handshake and 1-RTT packets (\texttt{step} 8b), forwarding them to the host only after confirming that the \texttt{DCID} is present in its records (\texttt{step} 9b).
Finally, all bidirectional data transmission 1-RTT packets are managed by the DPA cache (\texttt{step} n and m).
\section{Handshake Floods Defense on DPU}\label{sec:dpu-offloading}

In this section, we first present stateless function offloading on the Arm agent, and then describe the DPA cache, which accelerates the data path by caching verified connections.


\subsection{Unified Stateless Offloading}\label{sec:token-design}
To efficiently offload stateless \retry processing to the DPU, \sysname unifies the stateless functions under a single AES-GCM pipeline and proposes a universal token design that simultaneously supports both \retry and 0-RTT address validation.

\textbf{Unifying stateless functions with AES-GCM.}
Existing QUIC implementations employ various algorithms (\eg AES or RSA)~\cite{quic-impl-list} to protect tokens.
This diversity complicates efficient offloading and leads to suboptimal utilization of the DPU's hardware accelerators.
To address this issue, \sysname standardizes all stateless operations, including token generation, validation, and integrity tag computation, on AES-GCM-128.

Our rationale is threefold.
First, AES-GCM is an AEAD algorithm that simultaneously ensures the confidentiality of the token structure (preventing structural analysis) and the integrity of the token (preventing forgery).
Second, unlike asymmetric algorithms (\eg RSA) which impose prohibitive latency, AES-GCM is symmetric and highly efficient.
Third, the target BF3 DPU features a dedicated AES-GCM hardware engine. In practice, hardware-accelerated AES-GCM on the DPU incurs a lower per-packet latency (<2~$\mu$s) than host-side lightweight HMAC~\cite{hmac} computation ($\approx$3~$\mu$s), making AES-GCM a more efficient choice for stateless retry processing.
By unifying the stateless functions with a single AEAD pipeline, \sysname maximizes the utilization of the hardware accelerator and sustains high validation performance even under flood attacks.

\begin{figure}[htbp]
    \centering
    \includegraphics[width=\linewidth]{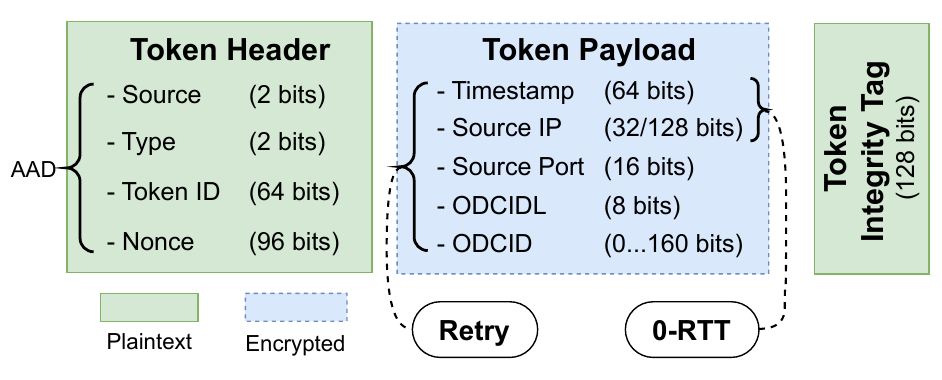}
    \caption{\newtext{Universal secure token format. The token header is authenticated as AAD when computing the integrity tag.}}
    \label{fig:token-format}
\end{figure}
\textbf{Universal token format.}
We introduce a universal token format (\fig~\ref{fig:token-format}), which supports both \retry and 0-RTT address validation.
Each token consists of a plaintext header, an AES-GCM-encrypted payload, and a 128-bit token integrity tag.
The header includes (i) a 2-bit source field indicating the token issuer (\texttt{0x1} for host and \texttt{0x2} for DPU), (ii) a 2-bit type field specifying the token purpose (\texttt{0x1} for \retry and \texttt{0x2} for 0-RTT), (iii) a token identifier (ID), and (iv) a 96-bit nonce.
The token header is authenticated as associated data (AAD) to bind the token metadata to the ciphertext.
The encrypted payload carries the information needed for stateless client validation and freshness checks.
For \retry tokens, the payload encodes a creation timestamp, the client's source IP address (32 bits for IPv4 and 128 bits for IPv6) and source port, and the original DCID (length and bytes) to bind the token to the intended connection context.
For 0-RTT tokens carried in \texttt{NEW\_TOKEN}, the payload similarly includes a timestamp and the client address tuple (omits the source port, ODCIDL, and ODCID), enabling the DPU to perform address validation for 0-RTT attempts without maintaining per-connection state.
Tokens that fail AEAD authentication or violate freshness/address checks are rejected.
\newtext{
Appendix Table~\ref{apptab:quic-token-summary} compares token formats and protection algorithms across QUIC implementations listed in~\cite{quic-impl-list}.
AES-GCM is the most commonly adopted protection mechanism in this set.
However, AES-GCM-based designs still differ substantially in their concrete token formats.
\sysname distills the common requirements of existing token designs---freshness, client-address binding, connection-context binding, and integrity protection---into a unified AES-GCM-128 token format.
This unified format allows \retry tokens and \texttt{NEW\_TOKEN} tokens to share the same DPU-side parsing and verification pipeline, making token validation hardware-friendly while preserving the security properties expected from existing QUIC token designs.
}

\textbf{0-RTT validation on the DPU.}
0-RTT reduces connection setup latency by allowing early data, but it also expands the replay surface~\cite{rfc8446}.
In QUIC, servers can issue address validation tokens to clients via \texttt{NEW\_TOKEN} frames for future connections~\cite{rfc9000}.
\sysname reuses the same universal token format for \texttt{NEW\_TOKEN} (type=\texttt{0x2}), so the DPU can validate 0-RTT attempts using the same stateless AES-GCM pipeline as \retry.
This design allows the DPU to independently gate 0-RTT handshakes based on authenticated token contents (including freshness and address binding), while keeping validation stateless.
We configure the 0-RTT token validity window to match the session ticket lifetime (\eg 24 hours), consistent with common TLS resumption deployments~\cite{rfc5246}.

\textbf{Lightweight key derivation with a single \newtext{pre-shared master key}.}
Offloading requires the host and the DPU to derive identical cryptographic materials without complex synchronization.
\sysname uses HKDF~\cite{hkdf} to derive per-connection AEAD keys from a single \newtext{pre-shared master key, $K_{\mathrm{psk}}$}, which is shared between the host and the DPU, periodically rotated, and shown in Algorithm~\ref{algo:key-gen}.
For each token, \sysname deterministically derives an initial secret using \texttt{HKDF-Extract} over $(K_{\mathrm{psk}},\, TID)$, where $TID$ is the randomly generated token ID.
It then derives a token secret $S_{\mathrm{token}}$ using \texttt{HKDF-Expand-Label} with label \texttt{retry} \texttt{token}, and finally expands $S_{\mathrm{token}}$ into an AES-GCM key $K_{\mathrm{enc}}$ and a base nonce $N_{\mathrm{base}}$ using distinct labels.
The token header is used as AAD, and the payload is encrypted to produce the integrity tag.

To avoid nonce reuse when multiple tokens are generated for the same $TID$ under the same $K_{\mathrm{psk}}$ (e.g., repeated \retry issuance), we derive the final per-token nonce by mixing runtime entropy into the base nonce.
Specifically, we compute $N_{\mathrm{enc}} = N_{\mathrm{base}} \oplus (T \parallel R)$, where $T$ is a 64-bit timestamp and $R$ is a 32-bit random value.
This construction ensures that each AEAD invocation uses a distinct nonce even under repeated token issuance.

\begin{algorithm}[htbp]
\caption{Derivation of the key and nonce}
\label{algo:key-gen}
\KwIn{\newtext{Pre-shared master key $K_{\mathrm{psk}}$} and token identifier $TID$}
\KwOut{Encryption key $K_{\mathrm{enc}}$ and nonce $N_{\mathrm{enc}}$}
\tcp{Denote HKDF-Extract as Extract}
\tcp{Denote HKDF-Expand-Label as Expand}
$T \gets \text{current\_timestamp}$ \tcp*{64-bit timestamp}
$R \gets \text{random\_bytes}(4)$ \tcp*{32-bit random value}
$S_{\mathrm{initial}} \gets \mathrm{Extract}(K_{\mathrm{psk}},\ TID)$\;
$S_{\mathrm{token}} \gets \mathrm{Expand}(S_{\mathrm{initial}},\ \texttt{"retry token"},\ \texttt{""},\ 32)$\;
$K_{\mathrm{enc}} \gets \mathrm{Expand}(S_{\mathrm{token}},\ \texttt{"token key"},\ \texttt{""},\ 16)$\;
$N_{\mathrm{enc}} \gets \mathrm{Expand}(S_{\mathrm{token}},\ \texttt{"token iv"},\ \texttt{""},\ 12)$\;
$N_{\mathrm{enc}} \gets N_{\mathrm{enc}} \oplus \ (T \parallel R)$ \;
\Return $(K_{\mathrm{enc}},\ N_{\mathrm{enc}})$\;
\end{algorithm}
\newtext{\textbf{Master key rotation.}
To limit the impact of potential key compromise, the pre-shared master key $K_{\mathrm{psk}}$ is rotated periodically, following the 0-RTT token expiration time, e.g., every 24 hours.
Because per-token cryptographic materials are derived from $K_{\mathrm{psk}}$ and diversified by $TID$, compromising one derived token key does not reveal the master key or the keys of other tokens.
Therefore, \sysname can adopt a relatively long rotation interval without maintaining per-connection keys.
To avoid dropping legitimate packets during rotation, tokens encrypted under the previous master key remain valid within a short transition window.}

\subsection{DPA Accelerated Data Transmission}\label{subsec:dpa-cache}
This section focuses on optimizing the data path to reduce data transmission latency. We first contrast our on-path processing approach with off-path designs that incur additional overhead, and then present the mechanisms employed by \sysname to reduce end-to-end transmission latency.

Offloading \retry to the off-path Arm processor on the DPU can significantly improve throughput.
However, it introduces additional latency for each packet. This overhead stems from the off-path architecture of the Arm processor, where all the incoming and outgoing packets must first be forwarded through the embedded PCIe switch (\textasciitilde 500 ns delay in BF3~\cite{demystifying-bf3}) before reaching the host. This traffic path is commonly referred to as \textit{slow path}.

An alternative traffic path powered by the on-path DPA processor, \ie the \textit{fast path}, can process all traffic and forward it directly to the host without passing the off-path Arm processor.
As discussed in \sect~\ref{subsec:bf3-arch}, the DPA processor sits on the critical path of traffic, and provides per-packet programmability for all traffic traversing it. 
That said, some \retry functions can be further offloaded from the Arm processor to the DPA processor to reduce the end-to-end transmission latency.

\textbf{Fast-path authorization.}
The three core Retry functions are ill-suited for the DPA fast path, as each relies on AES-GCM operations through the hardware accelerator attached to the Arm processor. 
Accessing this accelerator via the PCI switch adds extra latency,
thereby contradicting our objective of minimizing delay.
Conversely, \textit{connection authorization}, which involves only simple matching of DCIDs, is well suited for fast-path offloading.
In the offloading design, the fast path is dedicated to processing all packets except Initial packets, as Initial packets are handled by the Arm processor for stateless \retry.
Upon packet arrival, the DPA inspects the DCID field in the QUIC header to determine whether the connection ID belongs to the validated set. Packets with DCIDs not in the set are dropped, while those with valid DCIDs are forwarded directly to the host, bypassing the slow path.

\newtext{
\textbf{Bloom filters for approximate connection tracking.}
In practical applications, a server may handle tens of thousands of legitimate concurrent connections per second. Maintaining the DCID of every connection at the DPA would be prohibitively expensive in terms of hardware cost. To address this issue, the DPA cache introduces Bloom filters~\cite{bloomfilter} to track verified legitimate connections with extremely low memory overhead. The core advantage of this data structure lies in its absence of false negatives, which fundamentally ensures that benign connections will not be mistakenly blocked due to matching failures. Although Bloom filters have a very small probability of false positives, which may allow a small number of unverified packets to bypass the front-end check, this is an acceptable trade-off in the overall defense architecture.

\textbf{Aging mechanism based on time-window rotation.}
To prevent the false positive rate from deteriorating due to the long-term accumulation of stale states in the Bloom filter, the DPA cache adopts an automatic aging mechanism based on dual-table rotation over a time window ($T_W$). The DPA cache maintains two filters in parallel, periodically alternating them between the \texttt{ACTIVE} and \texttt{PASSIVE\_AGING} states. The liveness of a connection is dynamically refreshed by actual traffic: when a 1-RTT packet arrives, the DPA cache concurrently matches it against both tables; if it hits the passive aging table, the corresponding DCID is reinserted into the active table to extend its lifetime. This means that as long as continuous data exchange exists, the connection can migrate smoothly between the two tables without being cleared. In this mechanism, the choice of $T_W$ must carefully balance preventing idle connections from expiring prematurely against suppressing the increase in the filter’s false positive rate. Accordingly, the DPA cache sets $T_W$ to 30 seconds, following Chromium’s default maximum idle timeout to balance defense accuracy and throughput.

\textbf{Parameter trade-off of the Bloom filter.}
The Bloom filter in DPA cache uses $k=3$ hash functions and a total memory budget of $m=1$ MB. 
With an assumed arrival rate of $40$K QUIC connections per second and the rotation window $T_W=30$ seconds (mentioned before), the expected number of inserted elements is
$n = 40000 \times 30$. 
The resulting false positive rate (FPR) is
\begin{equation}
    FPR = \left(1 - e^{-\frac{k \cdot n}{m}}\right)^k \approx 4.25\%,
\end{equation}
where $m$ is measured in bits.

Note that this FPR does not weaken the protection against QUIC handshake floods because the Bloom filter is used only for fast authorization of post-handshake 1-RTT packets in the DPA cache.
Handshake-related packets are still handled by the Arm agent and are not admitted solely based on the Bloom filter result.
We further discuss attacks against the DPA cache in \sect~\ref{sec:security-analysis}.
}

\section{Host Cooperation}\label{sec:host-cooperation}
This section presents the design of the host-side cooperator.
\sysname redesigns the connection setup process to enable tight cooperation between the DPU and the host. This design preserves QUIC semantics for benign traffic and requires only minimal changes to the host QUIC stack.

While stateless \retry functions can be efficiently offloaded to the DPU (\sect~\ref{sec:dpu-offloading}), the stateful function, \emph{connection management}, is not amenable to offloading because it is tightly coupled with the host QUIC stack. 
\newtext{In particular, connection management is responsible for selecting, negotiating, and validating the CIDs used during the handshake. QUIC requires the relevant handshake CIDs, such as ODCID, RSCID, and the server's final SCID, to be encoded as transport parameters. These parameters are carried in the encrypted TLS Handshake payload and authenticated by the peer, as discussed in Section~\ref{subsec:handshake}. Since these CID values must remain consistent with the host's connection state and TLS handshake transcript, offloading connection management would require tight synchronization between the DPU and the host QUIC stack.}


\textbf{Strawman solution.} 
A simple approach is to run a QUIC proxy on the DPU's Arm processor.
In this design, the proxy completes the handshake with the client, establishes a separate QUIC connection with the host, and relays traffic between the two.
Although this approach preserves host transparency, it introduces additional latency during connection setup and increases system complexity by requiring connection state management on the Arm processor.

\textbf{Token-based connection state synchronization.}
\sysname instead adopts a lightweight cooperation mechanism that uses the Retry token as a carrier for connection identifier state.
The retry token format described in \sect~\ref{sec:token-design} encodes the ODCID within its encrypted payload.
After validating the token, the DPU forwards the Initial packet directly to the host without modification (mentioned in \sect~\ref{subsec:e2e-workflow}).
Upon receiving the validated Initial packet, the host derives the token decryption key using the \newtext{pre-shared master key ($K_{\mathrm{psk}}$)} and the TID extracted from the token, following Algorithm~\ref{algo:key-gen}.
It then decrypts the token payload and recovers the ODCID. Note that the RSCID (\texttt{S2}) is obtained directly from the DCID field of the received Initial packet.
Both identifiers are then passed to the QUIC stack, ensuring correct handshake processing and transport parameter construction.
Because the host and the DPU share only a single rotating \newtext{master key}, this design avoids the need to synchronize multiple cryptographic materials, which makes \sysname easy to deploy. 

\textbf{Time consistency.}
Maintaining clock consistency between the DPU and the host is essential. Any time drift in the DPU may lead to incorrect token validity checks, posing a risk to system security. Therefore, periodic time synchronization via NTP is necessary.

\textbf{Fail-open mechanism.}
\sysname is designed to operate in a fail-open manner, ensuring that DPU-side failures do not compromise security or service availability.
When the DPU-side program crashes or the Arm operating system on the DPU becomes unavailable, packet forwarding continues to be handled by the embedded switch on the NIC, allowing traffic to be transparently delivered to the host.
In this case, \sysname automatically falls back to a host-only processing path without interrupting existing or new connections.
The host QUIC stack implements the same stateless \retry logic as the DPU-offloaded path, allowing independent \retry processing even when the DPU is unavailable.
While this fallback path incurs higher CPU overhead and reduced performance compared to DPU acceleration, it preserves service availability and does not weaken security guarantees.

\section{Security Analysis}\label{sec:security-analysis}
We discuss possible attacks from adversaries against our system itself (above and beyond the QUIC handshake flooding attack vector), and explain how we address them.

\textbf{Token replay attacks.}
\sysname embeds the current UNIX timestamp into the token body and encrypts it using the key–nonce pair generated by Algorithm~\ref{algo:key-gen}.
When a QUIC Initial packet containing a token arrives, \sysname decrypts the token to extract the timestamp and checks whether it has expired.
The token expiration is set to 1 second, representing the common 1-RTT duration in the modern Internet.
This design ensures that \sysname accepts only retry tokens issued within a 1-RTT period, providing strong resilience against replay attacks with outdated tokens.

\textbf{Token forgery attacks.}
An adversary may send crafted QUIC Initial packets to \sysname and analyze the tokens contained in the returned Retry packets in an attempt to reverse-engineer the token generation mechanism.
If successful, the attacker could forge valid tokens and bypass \sysname. \textit{That said}, this process constitutes a Chosen-Plaintext Attack (CPA) against our token generation mechanism.
The security against token forgery directly reduces to the security of the AES-GCM encryption algorithm. AES-GCM is proven to be secure in the CPA model, provided that the input nonce is never reused with the same key~\cite{aes-gcm-security, aes-gcm-tip}.
In \sysname, for each key, the nonce is computed by performing an XOR operation with the timestamp (Algorithm~\ref{algo:key-gen}), ensuring nonce uniqueness for each key.
The proposed key–nonce derivation scheme provides security against token forgery attacks under the security guarantees of the AES-GCM algorithm.

\newtext{\textbf{DPU-targeted DDoS attacks.}
Since \sysname offloads \retry processing from the host CPU to the DPU, an adversary may attempt to attack the DPU itself by sending high-rate Initial packets, 1-RTT packets, or traffic patterns that intentionally trigger the DPU processing.
Such attacks aim to exhaust DPU Arm cores, crypto engines, and the DPA cache.
\sysname mitigates these attacks by keeping the DPU processing path bounded and stateless for unvalidated clients.
Malformed 1-RTT packets that fail DPA cache authorization are dropped in the fast path before reaching the Arm cores or the host.
When the DPU approaches saturation, the system can shed load at the DPU by dropping or rate-limiting unauthenticated Initial packets, rather than forwarding all traffic to the host and exposing the QUIC stack to the original flooding attack.
Thus, DPU saturation may degrade availability for new handshakes, but will not amplify the attack into host CPU exhaustion.

\textbf{State exhaustion attacks on the DPA cache.}
An adversary with many real source addresses may try to pollute the DPA cache by completing \retry for many short-lived connections. This can increase the Bloom filter false-positive rate within the current aging window. However, such an attack requires real address validation and client-side resources, making it substantially more expensive than spoofed Initial floods. The time-window-based aging mechanism bounds the duration of pollution. Once the saturated filter is retired, stale DCIDs are removed and the false-positive rate is restored. Thus, cache pollution may temporarily admit a bounded fraction of unauthorized DCIDs, but it does not permanently compromise the cache's ability to distinguish validated DCIDs from invalid ones.
}


\section{Evaluation}\label{sec:evaluation}

We evaluate \sysname by running several experiments on a hardware testbed.
We (1) demonstrate the throughput of \sysname in mitigating QUIC handshake floods, and (2) show the overall performance of \sysname in comparison to the host-based solution.
Note that we own all testbed infrastructure, and attack traffic was only directed to dedicated testbed servers, raising no ethical issues.

\subsection{Experimental Setup}\label{subsec:expr-setup}

\textbf{Prototype implementation.}
We implement \sysname on the BlueField-3 DPU~\cite{bf3} using DOCA v2.7.0~\cite{doca}, leveraging the DPA, the Arm processor, and AES-GCM hardware accelerators. The implementation comprises about 15,000 lines of code (LOC) in C/C++. For the host cooperation, we utilize both Cloudflare's quiche~\cite{quiche} (Rust) and aioquic~\cite{aioquic} (Python 3) libraries to implement the proposed collaborative retry mechanism with about 1,000 LOC.

\textbf{Testbed.} 
The testbed consists of three servers and an Intel Tofino Wedge32X-BF programmable switch serving as an L2 switch.
The server machines are all connected to the switch via Direct Attach Copper (DAC) cables and consistently run Ubuntu 22.04 with kernel version 5.15.0.
One server acts as the adversary, with a 48-core Intel Xeon Silver 4214R CPU @ 2.4Ghz and a Mellanox ConnectX-6 2$\times$100Gbps NIC, generating attack traffic using DPDK 23.03~\cite{dpdk}.
The second server functions as the client, featuring a 48-core Intel Xeon 4214R CPU@2.4Ghz and an Intel X710 2$\times$10Gbps NIC to send benign requests. 
The third server hosts the HTTP/3 service, powered by a 96-core Intel Xeon Gold 5318Y CPU@2.10 GHz and a BlueField-3 2$\times$200Gbps NIC~\cite{bf3}.

\textbf{Background traffic.}
We use real HTTP/3 requests as background traffic. An HTTP/3 client based on \newtext{quic-go~\cite{quic-go}} is implemented and deployed on the client machine. During the experiment, the client sends \newtext{40k} requests per second to the server running an HTTP/3 service and records both the connection setup latency and the response latency for each request. To prevent requests from hanging due to DDoS interference, the timeout for each request is set to 3 seconds.

\textbf{Attack traffic generation.}
We implement a QUIC handshake flood generator on the adversary machine with DPDK~\cite{dpdk}. It can generate 100Gbps attack traffic using the Mellanox ConnectX-6 NIC. The attack traffic primarily consists of QUIC Initial packets whose source IP (drawn from public prefixes), source port, DCID, and SCID are randomized.
The QUIC payload is sampled from legitimate QUIC Initial packets. Each attack packet is 1242 bytes in length, complying with the QUIC design requirement that the UDP payload in QUIC Initial packets must be at least 1200 bytes~\cite{rfc9000}.

\textbf{Evaluation metrics.}
We utilize the connection establishment latency (ms), the data transmission latency (ms) of QUIC, and the maximum sustainable attack rate without packet loss as key metrics to evaluate the system performance. Note that the attack rate metric is measured in kilo packets per second (kpps). Additionally, we measure the host CPU load using the Million Instructions Per Second (MIPS) metric to assess the effectiveness of \retry offloading.


\textbf{Baselines.}
\newtext{We implement three variants of our system. On the BF3 DPU, we implement a slow-path-only version (TurboRetry$^{\dagger}$) and a hybrid version (\sysname) that leverages both the slow-path and fast-path processors to reduce data transmission latency. To isolate the impact of the retry mechanism from that of hardware acceleration, we also implement a host-based \sysname variant, \ie \sysname-XDP, that ports the DPU-side functionality to the server host. Specifically, the Arm agent logic is implemented using \texttt{AF\_XDP}, where packets are steered to user-space retry workers for token generation and validation. The DPA cache is implemented in eBPF to provide fast in-kernel filtering. We compare these three variants with the end-host protocol stacks.}

\subsection{Resilience against Handshake Floods}

\newtext{Under the setup described above, we evaluate \sysname against host-based implementations (Aioquic~\cite{aioquic} and Quiche~\cite{quiche}) and \sysname-XDP.} Since the \retry in \sysname is primarily offloaded to the 16-core Arm processor, we further evaluate the speedup ratio with respect to the number of active cores. \newtext{For multi-core measurements of the protocol stack, we enable \texttt{SO\_REUSEPORT} so that multiple HTTP/3 worker processes listen on the same UDP port. We further attach an eBPF reuseport selector that dispatches incoming packets to workers based on the DCID, ensuring that packets belonging to the same connection are consistently delivered to the same worker. In addition, for each experiment, we configure IRQ affinity according to the number of active workers: the NIC receive queues are pinned to the CPU cores running the corresponding workers, reducing cross-core packet handoff and improving measurement stability.}

\textbf{Maximum sustained attack rate measurement.}
We measure the maximum attack rate, in Mpps (million packets per second), that each solution can sustain \textit{without incurring packet loss}. Each measurement is repeated 10 times to ensure that no packet loss occurs during packet processing. 

\begin{figure}[tp]
    \centering
\includegraphics[width=0.65\linewidth]{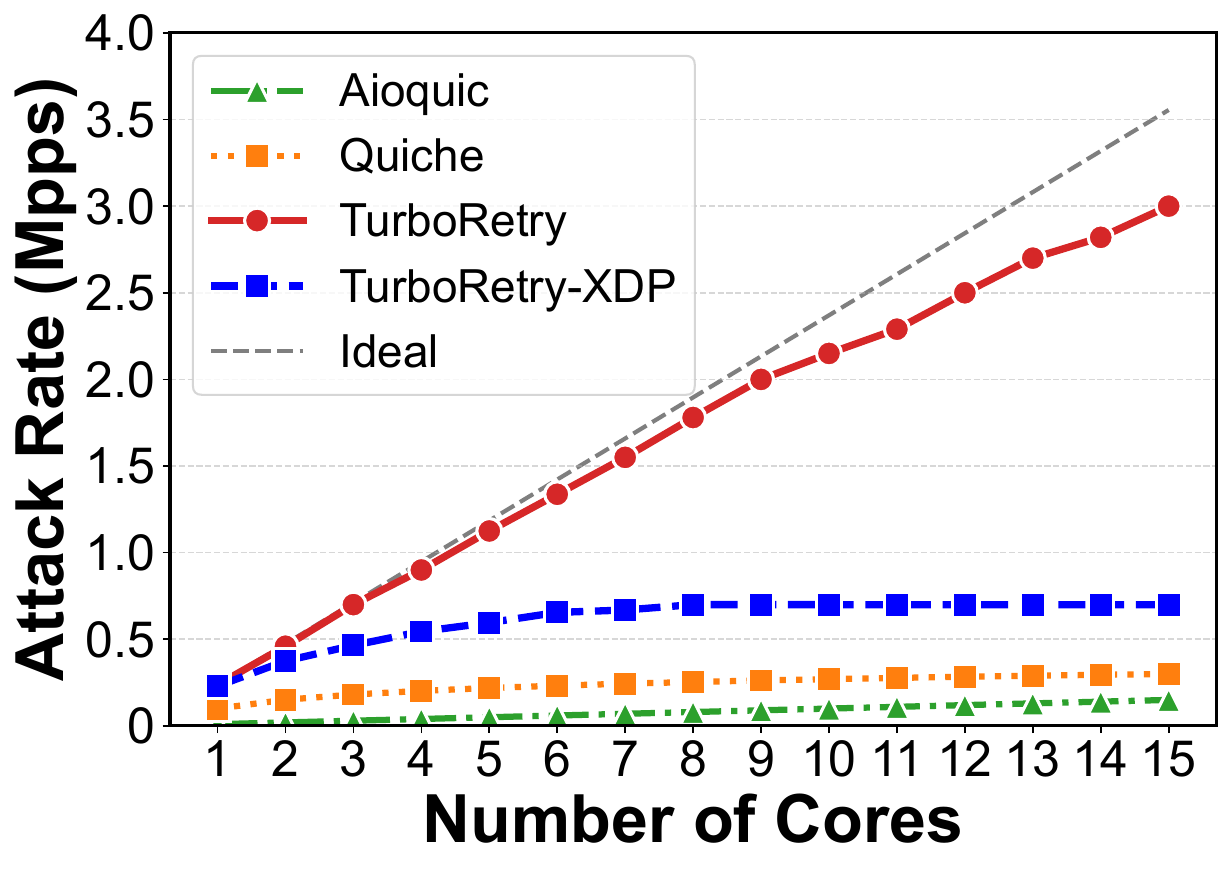}
    \caption{\newtext{Maximum sustained attack rate. \sysname defends against attacks \textit{without packet loss} until a rate of 3~Mpps, outperforming Aioquic and Quiche by $10\times$ and $20\times$, respectively.}}
    \label{fig:speedup-ratio}
\end{figure}

\textbf{Results.}
\newtext{As shown in \fig~\ref{fig:speedup-ratio}, \sysname outperforms the XDP-based variant (\sysname-XDP), Quiche, and Aioquic by $4.2\times$, $10\times$, and $20\times$, respectively, achieving the maximum sustained attack rate of 3~Mpps on the BlueField-3 DPU with 15 cores (1 left for the main thread).}
Importantly, this performance is achieved on the DPU’s embedded Arm processor cores, which operate at lower clock frequencies and provide substantially lower per-core compute capability than the x86 server-class CPUs used by Aioquic and Quiche.
Despite this hardware disadvantage, \sysname is able to sustain significantly higher attack rates, highlighting the effectiveness of offloading \retry processing to the DPU.
We further observe that 3~Mpps represents the maximum sustained attack rate of \sysname. When the attack rate exceeds this threshold, packet loss begins to occur, indicating that the system has reached its processing capacity.
The scalability trend shows that \sysname's maximum sustainable attack rate increases almost linearly with the number of parallel processor cores when fewer than nine cores are used.
Beyond this point, the growth rate tapers off, limiting further performance gains.
When the number of active cores is small (\eg, $\le 9$), the overhead of hardware-accelerated AES-GCM encryption is very low ($<2\mu s$) and accounts for only a small fraction of the total \retry processing time; thus, accelerator contention is negligible and the system scales nearly linearly.
As more cores process packets in parallel and the aggregate packet rate increases, accesses to the shared crypto accelerator become more frequent, causing contention and slowing down the increase in sustainable attack rate.
\newtext{In contrast, both the host-based and \sysname-XDP implementations do not achieve linear throughput scaling as the number of workers increases.
For the host-stack baselines, \texttt{SO\_REUSEPORT} only improves packet distribution across workers; it does not remove the shared bottlenecks in the host receive path.
Packets still traverse the NIC driver, skb allocation, UDP socket lookup, and reuseport selection before reaching the selected worker.
As the worker count increases, these shared per-packet costs, together with cache and memory-system contention, eventually dominate the execution time, causing throughput to plateau.
A similar effect appears in \sysname-XDP.
Although \texttt{AF\_XDP} bypasses part of the conventional kernel networking stack, the retry logic is still executed on host CPU cores.
Each packet still incurs costs for XDP execution, \texttt{AF\_XDP} ring and buffer management, user-space packet processing, and two AES-GCM operations.
As more retry workers are added, the host CPU execution resources, LLC capacity, memory bandwidth, and cache-coherence traffic become increasingly contended.
Consequently, \sysname-XDP improves throughput only up to the point where these host-side resources become saturated, after which adding more cores brings little additional benefit.
}

\subsection{Connection Setup Latency}\label{subsec:conn-latency}
In this experiment, we measure the connection setup latency of benign HTTP/3 clients while the server is subjected to handshake flooding attacks of varying intensities.
Both benign and attack traffic are generated as described in \sect~\ref{subsec:expr-setup}.
We compare \slowsysname and \sysname against host-based implementations.
\definecolor{myred}{RGB}{180, 0, 0}
\definecolor{mygreen}{RGB}{0, 150, 0}
\begin{table}[tbp]
    \centering
    \caption{Connection setup latency under no-attack conditions. Both \slowsysname and \sysname introduce negligible connection setup latency overhead ($\approx0.2$ ms).}
    \label{tab:setup-latency}
    \small
    \begin{tabular}{cccc} 
    \toprule
    Server           & \multicolumn{3}{c}{Latency (ms)} \\
                     \cmidrule(l){2-4} 
    Types & Baseline & +\slowsysname   & +\sysname \\ \midrule
    Aioquic        & 31.77   & 32.02 \textcolor{mygreen}{(+0.25)} &  32.03 \textcolor{mygreen}{(+0.26)}\\
    Quiche         & 22.96   & 23.18 \textcolor{mygreen}{(+0.22)} &  23.22 \textcolor{mygreen}{(+0.24)}\\
    \bottomrule
    \end{tabular}
\end{table}

\textbf{Connection setup latency measurement.}
The QUIC connection setup latency refers to the duration from when the client sends its first Initial packet to when it receives the first 1-RTT packet. To measure this latency, we modified an aioquic-based HTTP/3 client to record the time $T_1$ when the client sends its first QUIC Initial packet, and the time $T_2$ when the handshake completion event callback is triggered. The connection setup latency is then obtained by computing $T_2-T_1$. For each attack rate, we repeat the experiment 10 times, reporting the average setup latency. Note that in all experiments, the server-side retry mechanism adopts the token format described in \sect~\ref{sec:token-design}, and parses RSCID and ODCID from the token after decryption to synchronize connection identifiers in the QUIC stack (as described in \sect~\ref{sec:host-cooperation}).

\begin{figure}[tp]
    \centering
    \includegraphics[width=0.66\linewidth]{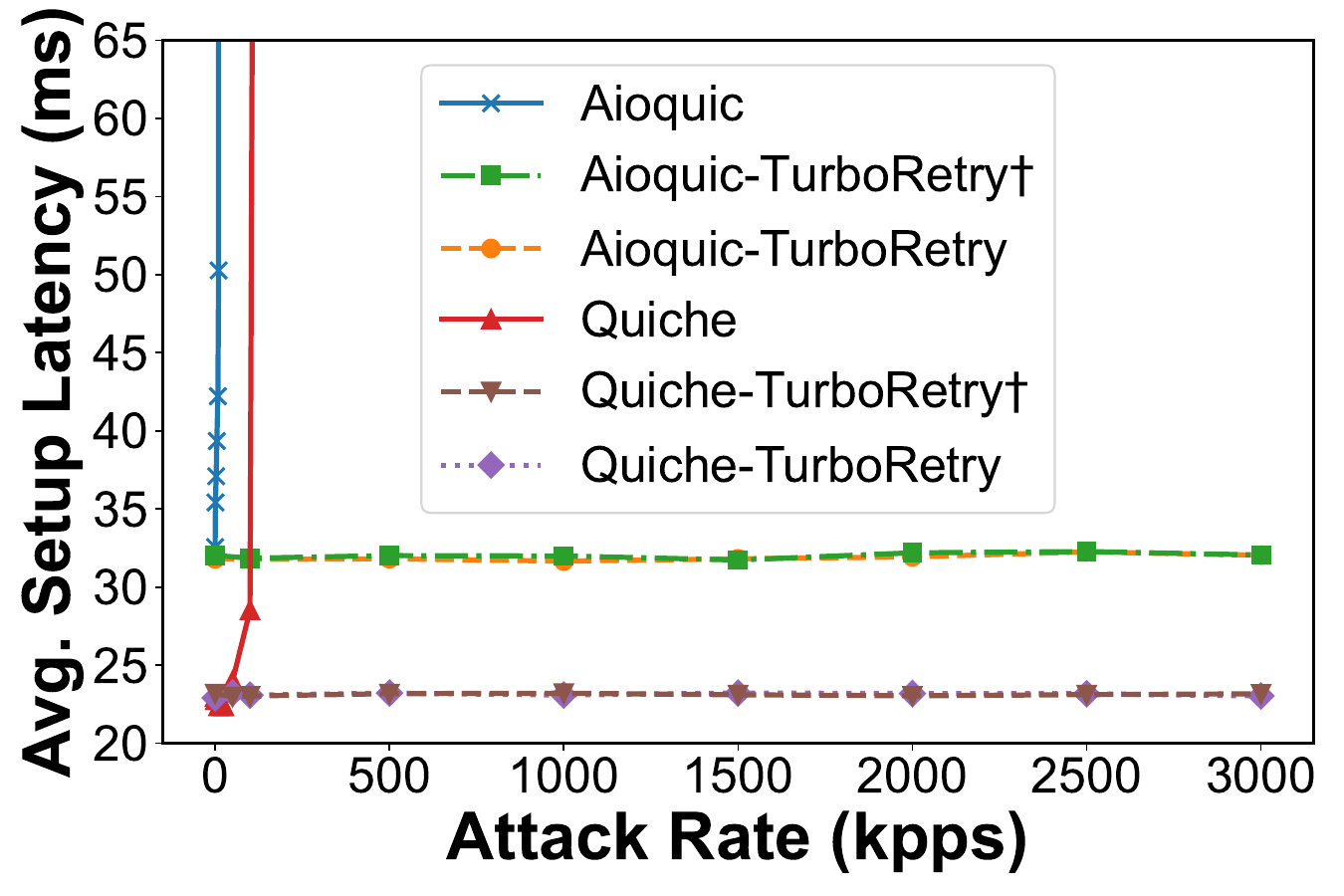}
    \caption{Connection setup latency under handshake flooding attacks. Both \slowsysname and \sysname maintain low connection setup latency under increasing attack rates with minimal overhead.}
    \label{fig:conn-latency}
\end{figure}

\begin{figure*}[!h]
    \centering

    \begin{subfigure}[t]{0.3\textwidth}
        \centering
        \includegraphics[width=\linewidth]{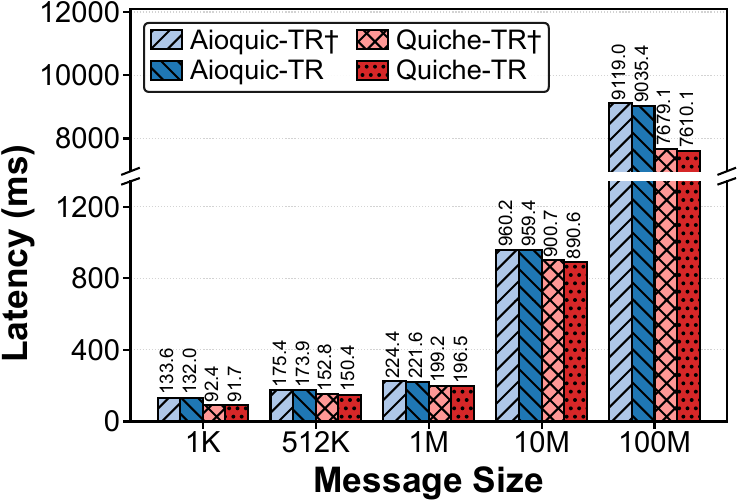}
        \caption{\newtext{No attack.}}
        \label{fig:latency-msg-size-0}
    \end{subfigure}
    \hfill
    \begin{subfigure}[t]{0.3\textwidth}
        \centering
        \includegraphics[width=\linewidth]{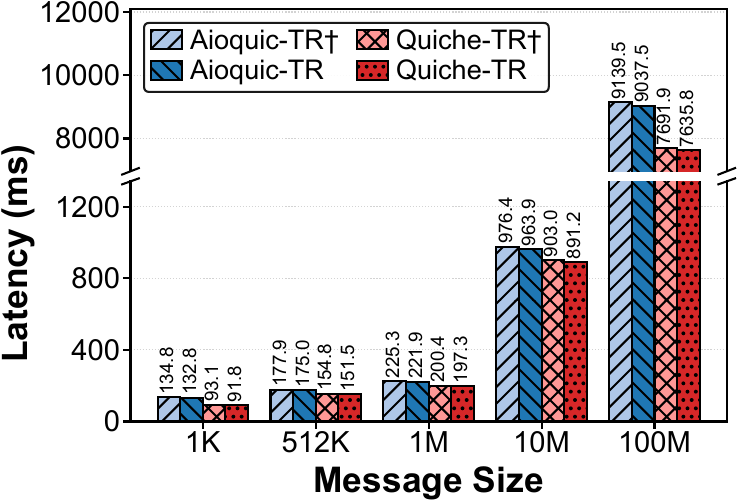}
        \caption{\newtext{1500~kpps attack.}}
        \label{fig:latency-msg-size-1500}
    \end{subfigure}
    \hfill
    \begin{subfigure}[t]{0.3\textwidth}
        \centering
        \includegraphics[width=\linewidth]{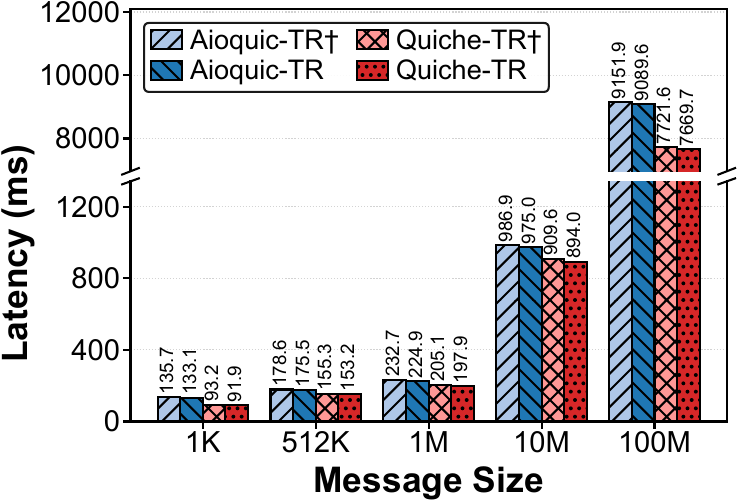}
        \caption{\newtext{3~Mpps attack.}}
        \label{fig:latency-msg-size-3000}
    \end{subfigure}

    \caption{\newtext{Transmission latency under different message sizes and attack rates.}}
    \label{fig:transmission-latency-msg-size}
\end{figure*}

\textbf{Results.}
As shown in Table~\ref{tab:setup-latency}, both \slowsysname (TR$^{\dagger}$) and \sysname (TR) introduce negligible connection setup overhead under no attack.
Across both Aioquic and Quiche, the additional latency remains within the sub-millisecond range ($\approx$0.2 ms).
As shown in \fig~\ref{fig:conn-latency}, servers equipped with \slowsysname and \sysname acceleration maintain low connection setup latency under varying attack intensities with minimal overhead. 
Both \slowsysname and \sysname exhibit similar connection setup latency overhead because the fast-path approach is primarily optimized for the data transfer stage. During the handshake, when the client receives the server’s 1-RTT packet, traffic in both cases still traverses the slow path.
However, in the original Aioquic~\cite{aioquic} and Quiche~\cite{quiche} implementations, connection establishment latency grows sharply as the attack rate increases, and both become unresponsive at approximately 20 kpps and 150 kpps, respectively.
It is worth noting that the difference in average connection establishment latency between Aioquic-based and Quiche-based solutions arises from differences in their host-side implementations.

\subsection{Data Transmission Latency}

In this section, we measure the data transmission latency of various file sizes (\eg 1 KB, 512 KB, 1 MB, 10 MB, and 100 MB). 
Following the same experimental setting in \sect~\ref{subsec:conn-latency}, we send HTTP/3 \texttt{GET} requests to the server under different attack rates, and measure the data transmission latency of these requests.
We then compare the performance of \slowsysname and \sysname with that of the host-based implementations.

\textbf{Data transmission latency measurement.}
The data transmission latency is defined by the time elapsed from connection setup to the connection closed of a \texttt{GET} request.
In particular, we record the whole request latency ($T_1$) and the connection setup latency ($T_2$), and obtain the data transmission latency as $T_1-T_2$. For each attack rate, we compute the average data transmission latency of all requests.
We use files of different sizes to emulate diverse workloads.
Files smaller than 1 MB (1 KB and 512 KB) represent typical web objects, such as HTML, JavaScript, CSS, and images. Files between 1 MB and 10 MB simulate compressed documents or short high-definition video clips. A large 100 MB file is used to emulate long video transfers, enabling a deeper evaluation of the stability and effectiveness of the offloaded retry mechanism. Note that attack rates are also measured in kilo packets per second (kpps).
\newtext{In addition, we use the normalized transmission latency overhead to quantify the overall extra transmission latency introduced by TurboRetry. Let $T_{host}$ denote the transmission latency of the host-based implementation and $T_{TR}$ denote the transmission latency of TurboRetry. This metric is defined as $(T_{TR} - T_{host}) / T_{host}$.}

\begin{table}[htbp]
    \caption{\newtext{The normalized transmission latency overhead.}}
    \label{tbl:additional-latency}
    \centering
    \small
    \begin{tabular}{ccc} 
    \toprule
    Server  & \multicolumn{2}{c}{Normalized Transmission Latency Overhead} \\
           \cmidrule(l){2-3} 
    Type & \slowsysname   & \sysname \\
    \midrule
    Aioquic & 2.0\% &  0.6\% \\
    Quiche  & 2.6\% &  1.1\% \\
    \midrule
    Average  & 2.3\% &  0.85\% \\
    \bottomrule
    \end{tabular}
\end{table}
\textbf{Results.}
\newtext{As shown in Figure~\ref{fig:transmission-latency-msg-size}, regardless of attack intensity, both \slowsysname and \sysname maintain consistent latencies across different file sizes. Compared with host-based implementations (Aioquic and Quiche), \slowsysname and \sysname introduce only about 2.3\% and 0.85\% additional transmission latency on average across various file sizes (shown in Table~\ref{tbl:additional-latency}), respectively.}
This shows that offloading data transmission from the slow path to the fast path further reduces the additional transmission latency.
Compared with \slowsysname, \sysname leverages the on-path RISC-V processor to handle validated connection traffic, eliminating the need to steer all packets to the Arm processor (slow path) and thereby significantly reducing NIC-to-host latency.

\subsection{Host CPU Load}

Using the same experimental setup as in \sect~\ref{subsec:expr-setup}, we measure the host CPU load under varying attack intensities. \sysname incurs no additional CPU load during QUIC handshake floods, reducing host CPU utilization by over 99\% compared with host-based implementations.
\newtext{Note that Aioquic can sustain only 8 kpps of concurrent connections on a single core. In this experiment, we configure Aioquic's background load as 8 kpps, while keeping Quiche’s background load at 40 kpps.}

\textbf{CPU load measurement.} 
We measure CPU load using \texttt{perf stat} to record instructions and cycles during HTTP/3 service.
With these values, we calculate MIPS (million instructions per second) to estimate CPU load.
Each measurement is repeated ten times, and the results are averaged. All CPU cores are fixed at 3.4GHz for more stable results.

\begin{figure}[htbp]
    \centering
    \includegraphics[width=0.68\linewidth]{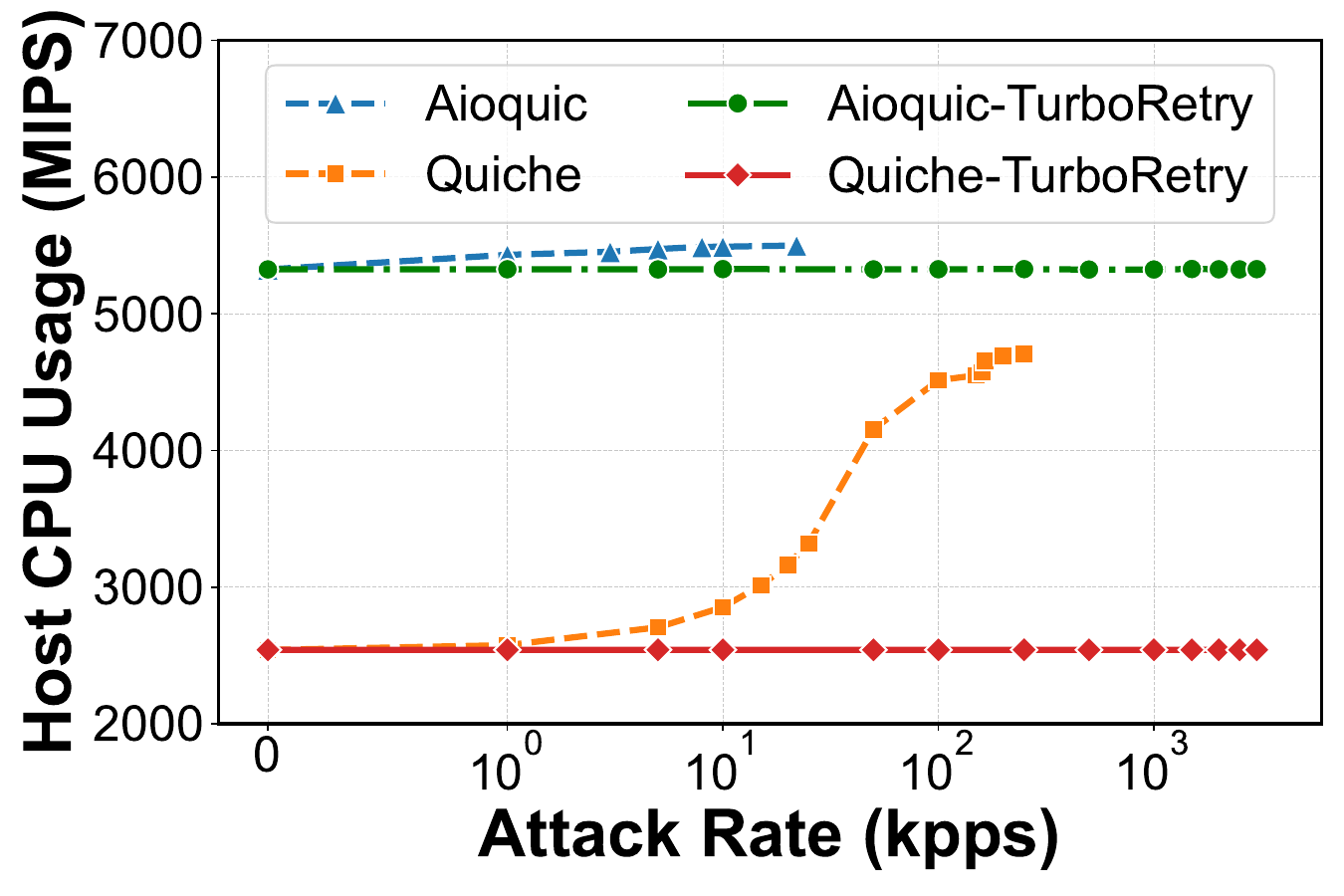}
    \caption{\newtext{Host CPU load. \sysname has no host CPU overhead against QUIC handshake flooding attacks.}}
    \label{fig:server-cpu-load}
\end{figure}
\textbf{Results.}
\fig~\ref{fig:server-cpu-load} shows that the CPU load of both Aioquic and Quiche rises sharply with increasing attack rates, peaking at 22 kpps and 150 kpps, respectively.
The difference in MIPS between Aioquic ($\approx 5400$) and Quiche ($\approx 4700$) stems from their execution efficiency. Across all measurements, Aioquic executes an average of 1.7 instructions per CPU cycle (IPC), whereas Quiche achieves 1.5 IPC.
Note that \sysname processes all QUIC handshake flooding packets entirely on the DPU, fully shielding the server CPUs from resource exhaustion during source address validation.
\newtext{With \sysname enabled, both Aioquic and Quiche maintain only the CPU utilization required for serving benign background traffic under a 3 Mpps attack, corresponding to 8 kpps for Aioquic ($\approx 5300$ MIPS) and 40 kpps for Quiche ($\approx 2500$ MIPS).}

\subsection{Fail-open Mechanism}
In this experiment, we evaluate the fail-open behavior of \sysname by examining the connection setup latency of benign clients when the DPU-side program becomes unavailable.
Fail-open ensures that system failures do not block service availability, and that traffic can continue to be served through the host path.

\textbf{Fail-open setup.}
To emulate a fail-stop failure, we explicitly terminate the DPU-side program at runtime while the client continuously generates benign HTTP/3 requests.
Throughout the experiment, the client issues requests at a fixed rate of 10 requests per second over a duration of 40 seconds.
At $t{=}\,20$s, we terminate the DPU program to trigger a failure.
When the DPU program becomes unavailable, packets are transparently processed by the host QUIC stack by design, allowing the system to fall back to the host path without interrupting service.

\begin{figure}[htbp]
    \centering
    \includegraphics[width=0.65\linewidth]{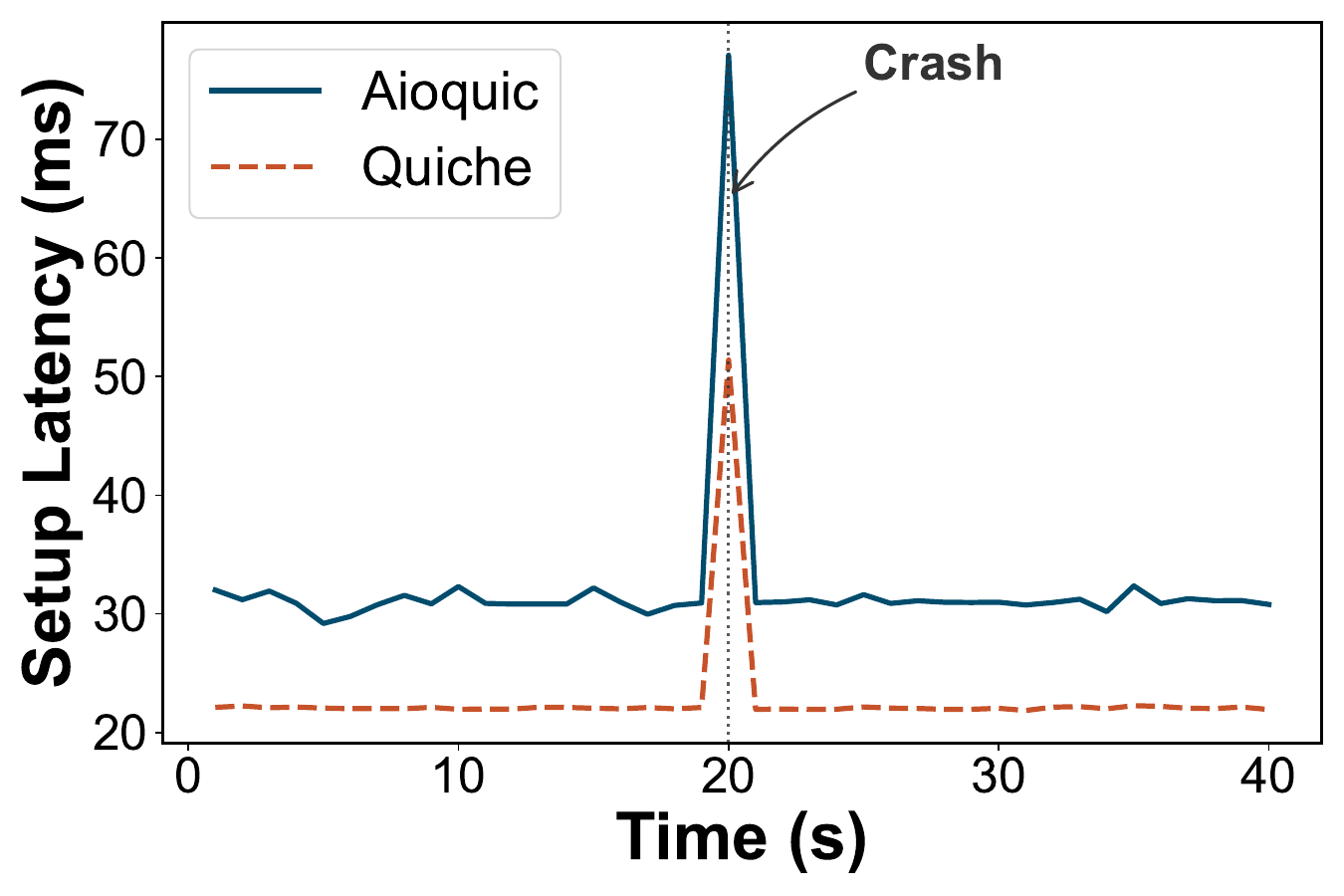}
    \caption{Connection setup latency during a fail-open event. When the DPU-side program crashes at $t{=}\,20$ s, a brief latency spike ($\approx30$ ms) is observed, reflecting the transition from DPU offloading to host-based processing.}
    \label{fig:fail-open}
\end{figure}
\textbf{Results.}
\fig~\ref{fig:fail-open} shows the connection setup latency when the DPU-side program is terminated at $t{=}\,20$ s.
Before the failure, \sysname maintains low and stable latency with DPU offloading enabled.
At the moment of termination, a brief latency spike of approximately 30~ms is observed, reflecting the transition from DPU offloading to host-based processing.
This spike is transient and bounded. Additionally, the latency converges to the corresponding host-only baseline and remains stable for the rest of the experiment.
No service disruption or request failures are observed, confirming that \sysname operates in a fail-open manner by transparently falling back to the host QUIC stack upon DPU failure.

\section{Discussion}\label{sec:discussion}

\textbf{Architectural restructuring of QUIC Retry.}
QFAM~\cite{qfam} proposes a retry-based defense mechanism to balance the resource usage between the attacker and the server.
It embeds a cryptographic challenge into the retry token and requires the client to solve it during the handshake process. 
Although this approach requires modifications to both client and server implementations, it addresses the inherent imbalance in resource consumption between attackers and victims during QUIC handshake flooding attacks. Our system, \sysname, can be integrated with QFAM~\cite{qfam} to further enhance its mitigation effectiveness.

\textbf{Extending to a protection cluster.}
The current evaluation of \sysname focuses on a single DPU-equipped server. Extending it to a protection cluster is a direction for future work.
In such a deployment, a group of servers equipped with DPUs could collaboratively mitigate handshake flooding attacks, allowing the aggregate defense capacity to scale with available host and DPU resources.
The split architecture of \sysname makes it well-suited for this extension, as mitigation logic can be scaled out without fundamental changes to the system design.

\textbf{Generality and applicability of \sysname.}
While our evaluation focuses on comparing HTTP/3 service performance across various QUIC implementations, \sysname can be extended to provide DDoS protection for other application-layer services built on QUIC, \eg DNS over QUIC~\cite{doq} (DoQ), gRPC over HTTP/3, and so on. This generality stems from the design of \sysname as a defense system operating at the QUIC protocol layer.

\newtext{
\textbf{Portability of \sysname across DPU vendors.} \sysname relies only on capabilities that are commonly available on modern DPUs, rather than vendor-specific primitives. The offloaded logic mainly consists of packet parsing, token generation, and token validation based on symmetric cryptography. Therefore, any DPU equipped with an AES-GCM acceleration engine can implement this functionality. The DPA cache of \sysname further relies on on-path packet steering and parsing capabilities, which allow the DPU to identify QUIC packets, extract the relevant header fields, and direct packets to the corresponding processing path. Such on-path steering support is widely provided by commercial DPUs for functions such as load balancing and security filtering.
}

\newtext{
\textbf{Deployment scenario.}
TurboRetry targets cloud servers where QUIC services run with DPU-assisted infrastructure offloading. 
Although host CPUs provide cryptographic instructions such as AES-NI, CPU-based Retry is not free: normal QUIC processing already consumes host cryptographic resources, and Retry further adds token generation, integrity tag computation, and token verification under attack traffic. 
This extra cryptographic data-plane workload can make host-based Retry a bottleneck under large-scale handshake floods. 
Moreover, in cloud environments, host CPU cores are tenant-facing resources. Recent SmartNIC/DPU systems similarly motivate offloading by preserving server resources for paying customers~\cite{humphries2025wave}, saving host CPU cycles for repetitive data-plane I/O~\cite{kim2023rearchitecting}, and avoiding contention between host-side infrastructure processing and tenant VMs~\cite{li2024triton}. 
Therefore, TurboRetry adopts a split design that offloads stateless Retry processing to the DPU-side AES-GCM pipeline and retains stateful connection management in the host QUIC stack.
}

\textbf{Latency evaluation of the host-based variant.} TurboRetry-XDP is evaluated only in terms of its maximum sustainable attack rate, and we do not report its connection setup or data transmission latency. Therefore, our latency results compare the DPU-based variants with the original Aioquic and Quiche stacks, rather than with a functionally equivalent host-based TurboRetry implementation. Evaluating the latency overhead of TurboRetry-XDP is left for future work.


\section{Related Work}\label{sec:related-work}

\textbf{Defense against QUIC handshake floods.}
Prior work can be classified into software-based and hardware-based schemes.
Software-based solutions leverage sophisticated algorithms to detect and mitigate QUIC handshake floods.
QUICShield~\cite{quicshield} detects malicious traffic with Bloom filters and utilizes the general similarity ratio CUSUM algorithm~\cite{sqcusum} to adapt to evolving attack patterns.
QUICwand~\cite{quicwand} further leverages dynamic counting Bloom filters~\cite{cbf} and a Bayesian optimizer~\cite{lightgbm, bayesian-opt1, bayesian-opt2} for automatic parameter tuning to reduce false positives.
QUICPro~\cite{quicpro} applies reinforcement learning for detection.
A hardware-based solution~\cite{mcia} uses Intel QuickAssist Technology~\cite{qat} to drop QUIC Initial packets with invalid integrity tags, but remains vulnerable to replay attacks.

\textbf{DPU-based TLS offloading.}
Previous studies~\cite{dpu-ktls, dpu-tcp-tls, iotcp, pno} have leveraged DPUs to offload cryptographic algorithms, thereby reducing the CPU burden on the host. However, these works primarily focus on enhancing the performance of the TCP stack or TLS over TCP, with limited exploration of QUIC offloading.

\textbf{In-network DDoS defense.}
Previous studies~\cite{accturbo,jaqen,lemon,smartcookie,poseidon,ripple} have employed programmable switches to mitigate a wide spectrum of DDoS attacks beyond QUIC handshake flooding. These approaches are complementary to \sysname and can be integrated to address other types of DDoS flooding attacks, \eg SYN flooding, UDP flooding, \etc, to further protect the server.

\section{Conclusion}\label{sec:conclusion}

The modern transport protocol QUIC is designed to enhance
network performance and security, but both QUIC and its built-in \retry defense remain vulnerable to handshake flooding attacks.
We design and implement a defense system, \sysname, with a split design that offloads the QUIC \retry mechanism onto off-the-shelf DPUs to mitigate large-scale QUIC handshake floods with high performance.
Through extensive experiments and analysis, we show that \sysname outperforms the host-side implementation by a wide margin, delivering a 10--20$\times$ throughput improvement while only introducing negligible additional latency.

\begin{acks}
We want to thank our anonymous reviewers for their valuable comments. This work is supported by the National Natural Science Foundation of China (Grant No. 62472404). 
\end{acks}

\bibliographystyle{ACM-Reference-Format}
\balance
\bibliography{arxiv}


\begin{thebibliography}{69}


\ifx \showCODEN    \undefined \def \showCODEN     #1{\unskip}     \fi
\ifx \showISBNx    \undefined \def \showISBNx     #1{\unskip}     \fi
\ifx \showISBNxiii \undefined \def \showISBNxiii  #1{\unskip}     \fi
\ifx \showISSN     \undefined \def \showISSN      #1{\unskip}     \fi
\ifx \showLCCN     \undefined \def \showLCCN      #1{\unskip}     \fi
\ifx \shownote     \undefined \def \shownote      #1{#1}          \fi
\ifx \showarticletitle \undefined \def \showarticletitle #1{#1}   \fi
\ifx \showURL      \undefined \def \showURL       {\relax}        \fi
\providecommand\bibfield[2]{#2}
\providecommand\bibinfo[2]{#2}
\providecommand\natexlab[1]{#1}
\providecommand\showeprint[2][]{arXiv:#2}

\bibitem[{aiortc orgnization}(2025)]%
        {aioquic}
\bibfield{author}{\bibinfo{person}{{aiortc orgnization}}.} \bibinfo{year}{2025}\natexlab{}.
\newblock \bibinfo{title}{{aioquic}}.
\newblock \bibinfo{howpublished}{https://github.com/aiortc/aioquic}.
\newblock


\bibitem[Alcoz et~al\mbox{.}(2022)]%
        {accturbo}
\bibfield{author}{\bibinfo{person}{Albert~Gran Alcoz}, \bibinfo{person}{Martin Strohmeier}, \bibinfo{person}{Vincent Lenders}, {and} \bibinfo{person}{Laurent Vanbever}.} \bibinfo{year}{2022}\natexlab{}.
\newblock \showarticletitle{Aggregate-based congestion control for pulse-wave DDoS defense}. In \bibinfo{booktitle}{\emph{Proceedings of the ACM SIGCOMM 2022 Conference}} (Amsterdam, Netherlands) \emph{(\bibinfo{series}{SIGCOMM '22})}. \bibinfo{publisher}{Association for Computing Machinery}, \bibinfo{address}{New York, NY, USA}, \bibinfo{pages}{693--706}.
\newblock
\showISBNx{9781450394208}
\href{https://doi.org/10.1145/3544216.3544263}{doi:\nolinkurl{10.1145/3544216.3544263}}


\bibitem[Antonakakis et~al\mbox{.}(2017)]%
        {marai}
\bibfield{author}{\bibinfo{person}{Manos Antonakakis}, \bibinfo{person}{Tim April}, \bibinfo{person}{Michael Bailey}, \bibinfo{person}{Matt Bernhard}, \bibinfo{person}{Elie Bursztein}, \bibinfo{person}{Jaime Cochran}, \bibinfo{person}{Zakir Durumeric}, \bibinfo{person}{J.~Alex Halderman}, \bibinfo{person}{Luca Invernizzi}, \bibinfo{person}{Michalis Kallitsis}, \bibinfo{person}{Deepak Kumar}, \bibinfo{person}{Chaz Lever}, \bibinfo{person}{Zane Ma}, \bibinfo{person}{Joshua Mason}, \bibinfo{person}{Damian Menscher}, \bibinfo{person}{Chad Seaman}, \bibinfo{person}{Nick Sullivan}, \bibinfo{person}{Kurt Thomas}, {and} \bibinfo{person}{Yi Zhou}.} \bibinfo{year}{2017}\natexlab{}.
\newblock \showarticletitle{{Understanding the Mirai Botnet}}. In \bibinfo{booktitle}{\emph{26th USENIX Security Symposium (USENIX Security 17)}}. \bibinfo{publisher}{USENIX Association}, \bibinfo{address}{Vancouver, BC}, \bibinfo{pages}{1093--1110}.
\newblock


\bibitem[Bishop(2022)]%
        {http3rfc}
\bibfield{author}{\bibinfo{person}{Mike Bishop}.} \bibinfo{year}{2022}\natexlab{}.
\newblock \bibinfo{title}{{HTTP/3}}.
\newblock \bibinfo{howpublished}{RFC 9114}.
\newblock
\href{https://doi.org/10.17487/RFC9114}{doi:\nolinkurl{10.17487/RFC9114}}


\bibitem[Bloom(1970)]%
        {bloomfilter}
\bibfield{author}{\bibinfo{person}{Burton~H Bloom}.} \bibinfo{year}{1970}\natexlab{}.
\newblock \showarticletitle{Space/time trade-offs in hash coding with allowable errors}.
\newblock \bibinfo{journal}{\emph{Commun. ACM}} \bibinfo{volume}{13}, \bibinfo{number}{7} (\bibinfo{year}{1970}), \bibinfo{pages}{422--426}.
\newblock


\bibitem[Chatzoglou et~al\mbox{.}(2022)]%
        {revisiting-quic-attacks}
\bibfield{author}{\bibinfo{person}{Efstratios Chatzoglou}, \bibinfo{person}{Vasileios Kouliaridis}, \bibinfo{person}{Georgios Karopoulos}, {and} \bibinfo{person}{Georgios Kambourakis}.} \bibinfo{year}{2022}\natexlab{}.
\newblock \showarticletitle{{Revisiting QUIC attacks: a comprehensive review on QUIC security and a hands-on study}}.
\newblock \bibinfo{journal}{\emph{Int. J. Inf. Secur.}} \bibinfo{volume}{22}, \bibinfo{number}{2} (\bibinfo{date}{Dec.} \bibinfo{year}{2022}), \bibinfo{pages}{347--365}.
\newblock
\showISSN{1615-5262}


\bibitem[Chen et~al\mbox{.}(2024a)]%
        {bf3-sketch}
\bibfield{author}{\bibinfo{person}{Xiang Chen}, \bibinfo{person}{Xi Sun}, \bibinfo{person}{Wenbin Zhang}, \bibinfo{person}{Xin Yao}, \bibinfo{person}{Zizheng Wang}, \bibinfo{person}{Hongyan Liu}, \bibinfo{person}{Qun Huang}, \bibinfo{person}{Gaoning Pan}, \bibinfo{person}{Xuan Liu}, \bibinfo{person}{Haifeng Zhou}, {and} \bibinfo{person}{Chunming Wu}.} \bibinfo{year}{2024}\natexlab{a}.
\newblock \showarticletitle{{Accelerating Sketch-based End-Host Traffic Measurement with Automatic DPU Offloading}}. In \bibinfo{booktitle}{\emph{IEEE INFOCOM 2024 - IEEE Conference on Computer Communications}}. \bibinfo{pages}{171--180}.
\newblock
\href{https://doi.org/10.1109/INFOCOM52122.2024.10621293}{doi:\nolinkurl{10.1109/INFOCOM52122.2024.10621293}}


\bibitem[Chen et~al\mbox{.}(2024b)]%
        {demystifying-bf3}
\bibfield{author}{\bibinfo{person}{Xuzheng Chen}, \bibinfo{person}{Jie Zhang}, \bibinfo{person}{Ting Fu}, \bibinfo{person}{Yifan Shen}, \bibinfo{person}{Shu Ma}, \bibinfo{person}{Kun Qian}, \bibinfo{person}{Lingjun Zhu}, \bibinfo{person}{Chao Shi}, \bibinfo{person}{Yin Zhang}, \bibinfo{person}{Ming Liu}, {and} \bibinfo{person}{Zeke Wang}.} \bibinfo{year}{2024}\natexlab{b}.
\newblock \showarticletitle{{Demystifying Datapath Accelerator Enhanced Off-path SmartNIC}}. In \bibinfo{booktitle}{\emph{2024 IEEE 32nd International Conference on Network Protocols (ICNP)}}. \bibinfo{pages}{1--12}.
\newblock


\bibitem[{Cloudflare, Inc}(2025)]%
        {quiche}
\bibfield{author}{\bibinfo{person}{{Cloudflare, Inc}}.} \bibinfo{year}{2025}\natexlab{}.
\newblock \bibinfo{title}{{quiche}}.
\newblock \bibinfo{howpublished}{https://github.com/cloudflare/quiche}.
\newblock


\bibitem[Cui et~al\mbox{.}(2022)]%
        {mcia}
\bibfield{author}{\bibinfo{person}{Bo Cui}, \bibinfo{person}{Zixuan Li}, {and} \bibinfo{person}{Fei Yu}.} \bibinfo{year}{2022}\natexlab{}.
\newblock \showarticletitle{{Manipulated Client Initial Attack and Defense of QUIC}}. In \bibinfo{booktitle}{\emph{2022 IEEE 24th Int Conf on High Performance Computing \& Communications; 8th Int Conf on Data Science \& Systems; 20th Int Conf on Smart City; 8th Int Conf on Dependability in Sensor, Cloud \& Big Data Systems \& Application (HPCC/DSS/SmartCity/DependSys)}}. \bibinfo{pages}{611--618}.
\newblock


\bibitem[Duke and Banks(2022)]%
        {ietf-quic-retry-offload}
\bibfield{author}{\bibinfo{person}{Martin Duke} {and} \bibinfo{person}{Nick Banks}.} \bibinfo{year}{2022}\natexlab{}.
\newblock \bibinfo{booktitle}{\emph{{QUIC Retry Offload}}}.
\newblock \bibinfo{type}{Internet-Draft} draft-ietf-quic-retry-offload-00. \bibinfo{institution}{Internet Engineering Task Force}.
\newblock
\shownote{Work in Progress}.
\newblock
\urldef\tempurl%
\url{https://datatracker.ietf.org/doc/draft-ietf-quic-retry-offload/00/}
\showURL{%
\tempurl}


\bibitem[Dworkin(2007)]%
        {aes-gcm-tip}
\bibfield{author}{\bibinfo{person}{Morris Dworkin}.} \bibinfo{year}{2007}\natexlab{}.
\newblock \bibinfo{title}{Recommendation for Block Cipher Modes of Operation: Galois/Counter Mode (GCM) and GMAC}.
\newblock
\urldef\tempurl%
\url{https://tsapps.nist.gov/publication/get_pdf.cfm?pub_id=51288}
\showURL{%
\tempurl}


\bibitem[Foundation(2015)]%
        {dpdk}
\bibfield{author}{\bibinfo{person}{Linux Foundation}.} \bibinfo{year}{2015}\natexlab{}.
\newblock \bibinfo{title}{{Data Plane Development Kit ({DPDK})}}.
\newblock
\urldef\tempurl%
\url{https://dpdk.org}
\showURL{%
\tempurl}


\bibitem[Gbur and Tschorsch(2023)]%
        {quicforge}
\bibfield{author}{\bibinfo{person}{Konrad~Yuri Gbur} {and} \bibinfo{person}{Florian Tschorsch}.} \bibinfo{year}{2023}\natexlab{}.
\newblock \showarticletitle{{QUICforge: Client-side Request Forgery in QUIC}}. In \bibinfo{booktitle}{\emph{30th Annual Network and Distributed System Security Symposium ({NDSS} 2023), San Diego, USA}}. \bibinfo{publisher}{The Internet Society}.
\newblock


\bibitem[Honda et~al\mbox{.}(2011)]%
        {tcp-middlebox-security}
\bibfield{author}{\bibinfo{person}{Michio Honda}, \bibinfo{person}{Yoshifumi Nishida}, \bibinfo{person}{Costin Raiciu}, \bibinfo{person}{Adam Greenhalgh}, \bibinfo{person}{Mark Handley}, {and} \bibinfo{person}{Hideyuki Tokuda}.} \bibinfo{year}{2011}\natexlab{}.
\newblock \showarticletitle{{Is it still possible to extend TCP?}}. In \bibinfo{booktitle}{\emph{Proceedings of the 2011 ACM SIGCOMM Conference on Internet Measurement Conference}} \emph{(\bibinfo{series}{IMC '11})}. \bibinfo{publisher}{Association for Computing Machinery}, \bibinfo{address}{New York, NY, USA}, \bibinfo{pages}{181--194}.
\newblock


\bibitem[Huitema et~al\mbox{.}(2022)]%
        {doq}
\bibfield{author}{\bibinfo{person}{Christian Huitema}, \bibinfo{person}{Sara Dickinson}, {and} \bibinfo{person}{Allison Mankin}.} \bibinfo{year}{2022}\natexlab{}.
\newblock \bibinfo{title}{{DNS over Dedicated QUIC Connections}}.
\newblock \bibinfo{howpublished}{RFC 9250}.
\newblock
\href{https://doi.org/10.17487/RFC9250}{doi:\nolinkurl{10.17487/RFC9250}}


\bibitem[Humphries et~al\mbox{.}(2025)]%
        {humphries2025wave}
\bibfield{author}{\bibinfo{person}{Jack~Tigar Humphries}, \bibinfo{person}{Neel Natu}, \bibinfo{person}{Kostis Kaffes}, \bibinfo{person}{Stanko Novakovi{\'c}}, \bibinfo{person}{Paul Turner}, \bibinfo{person}{Henry~M Levy}, \bibinfo{person}{David Culler}, {and} \bibinfo{person}{Christos Kozyrakis}.} \bibinfo{year}{2025}\natexlab{}.
\newblock \showarticletitle{Wave: Offloading resource management to SmartNIC cores}. In \bibinfo{booktitle}{\emph{Proceedings of the 30th ACM International Conference on Architectural Support for Programming Languages and Operating Systems, Volume 3}}. \bibinfo{pages}{264--281}.
\newblock


\bibitem[Injadat et~al\mbox{.}(2018)]%
        {bayesian-opt2}
\bibfield{author}{\bibinfo{person}{MohammadNoor Injadat}, \bibinfo{person}{Fadi Salo}, \bibinfo{person}{Ali~Bou Nassif}, \bibinfo{person}{Aleksander Essex}, {and} \bibinfo{person}{Abdallah Shami}.} \bibinfo{year}{2018}\natexlab{}.
\newblock \showarticletitle{{Bayesian Optimization with Machine Learning Algorithms Towards Anomaly Detection}}. In \bibinfo{booktitle}{\emph{2018 IEEE Global Communications Conference (GLOBECOM)}}. \bibinfo{pages}{1--6}.
\newblock
\href{https://doi.org/10.1109/GLOCOM.2018.8647714}{doi:\nolinkurl{10.1109/GLOCOM.2018.8647714}}


\bibitem[{Intel}(2025)]%
        {aes-ni}
\bibfield{author}{\bibinfo{person}{{Intel}}.} \bibinfo{year}{2025}\natexlab{}.
\newblock \bibinfo{title}{{Intel Advanced Encryption Standard Instructions (AES-NI)}}.
\newblock \bibinfo{howpublished}{https://www.intel.com/content/www/us/en/developer/articles/technical/advanced-encryption-standard-instructions-aes-ni.html}.
\newblock


\bibitem[Intel(2025)]%
        {qat}
\bibfield{author}{\bibinfo{person}{Intel}.} \bibinfo{year}{2025}\natexlab{}.
\newblock \bibinfo{title}{What Is Intel QuickAssist Technology (Intel QAT)}.
\newblock \bibinfo{howpublished}{\url{https://www.intel.com/content/www/us/en/products/docs/accelerator-engines/what-is-intel-qat.html}}.
\newblock


\bibitem[Iwata et~al\mbox{.}(2012)]%
        {aes-gcm-security}
\bibfield{author}{\bibinfo{person}{Tetsu Iwata}, \bibinfo{person}{Keisuke Ohashi}, {and} \bibinfo{person}{Kazuhiko Minematsu}.} \bibinfo{year}{2012}\natexlab{}.
\newblock \showarticletitle{Breaking and Repairing GCM Security Proofs}. In \bibinfo{booktitle}{\emph{Advances in Cryptology -- CRYPTO 2012}}, \bibfield{editor}{\bibinfo{person}{Reihaneh Safavi-Naini} {and} \bibinfo{person}{Ran Canetti}} (Eds.). \bibinfo{publisher}{Springer Berlin Heidelberg}, \bibinfo{address}{Berlin, Heidelberg}, \bibinfo{pages}{31--49}.
\newblock
\showISBNx{978-3-642-32009-5}


\bibitem[Iyengar and Thomson(2021)]%
        {rfc9000}
\bibfield{author}{\bibinfo{person}{Jana Iyengar} {and} \bibinfo{person}{Martin Thomson}.} \bibinfo{year}{2021}\natexlab{}.
\newblock \bibinfo{title}{{QUIC: A UDP-Based Multiplexed and Secure Transport}}.
\newblock \bibinfo{howpublished}{RFC 9000}.
\newblock
\href{https://doi.org/10.17487/RFC9000}{doi:\nolinkurl{10.17487/RFC9000}}


\bibitem[Jabbari et~al\mbox{.}(2024)]%
        {qfam}
\bibfield{author}{\bibinfo{person}{Abdollah Jabbari}, \bibinfo{person}{Y~A Joarder}, \bibinfo{person}{Benjamin Teyssier}, {and} \bibinfo{person}{Carol Fung}.} \bibinfo{year}{2024}\natexlab{}.
\newblock \bibinfo{title}{QFAM: Mitigating QUIC Handshake Flooding Attacks Through Crypto Challenges}.
\newblock
\showeprint[arxiv]{2412.08936}~[cs.CR]
\urldef\tempurl%
\url{https://arxiv.org/abs/2412.08936}
\showURL{%
\tempurl}


\bibitem[Joarder and Fung(2022)]%
        {quic-security-survey}
\bibfield{author}{\bibinfo{person}{Y~A Joarder} {and} \bibinfo{person}{Carol Fung}.} \bibinfo{year}{2022}\natexlab{}.
\newblock \showarticletitle{{A Survey on the Security Issues of QUIC}}. In \bibinfo{booktitle}{\emph{2022 6th Cyber Security in Networking Conference (CSNet)}}. \bibinfo{pages}{1--8}.
\newblock
\href{https://doi.org/10.1109/CSNet56116.2022.9955622}{doi:\nolinkurl{10.1109/CSNet56116.2022.9955622}}


\bibitem[Joarder and Fung(2024a)]%
        {quic-security}
\bibfield{author}{\bibinfo{person}{Y~A Joarder} {and} \bibinfo{person}{Carol Fung}.} \bibinfo{year}{2024}\natexlab{a}.
\newblock \showarticletitle{{Exploring QUIC Security and Privacy: A Comprehensive Survey on QUIC Security and Privacy Vulnerabilities, Threats, Attacks, and Future Research Directions}}.
\newblock \bibinfo{journal}{\emph{IEEE Transactions on Network and Service Management}} \bibinfo{volume}{21}, \bibinfo{number}{6} (\bibinfo{year}{2024}), \bibinfo{pages}{6953--6973}.
\newblock
\href{https://doi.org/10.1109/TNSM.2024.3457858}{doi:\nolinkurl{10.1109/TNSM.2024.3457858}}


\bibitem[Joarder and Fung(2024b)]%
        {quicpro}
\bibfield{author}{\bibinfo{person}{Y~A Joarder} {and} \bibinfo{person}{Carol Fung}.} \bibinfo{year}{2024}\natexlab{b}.
\newblock \showarticletitle{{QUICPro: Integrating Deep Reinforcement Learning to Defend against QUIC Handshake Flooding Attacks}}. In \bibinfo{booktitle}{\emph{Proceedings of the 2024 Applied Networking Research Workshop}} \emph{(\bibinfo{series}{ANRW '24})}. \bibinfo{publisher}{Association for Computing Machinery}, \bibinfo{address}{New York, NY, USA}, \bibinfo{pages}{94--96}.
\newblock


\bibitem[Joarder and Fung(2024c)]%
        {quicwand}
\bibfield{author}{\bibinfo{person}{Y~A Joarder} {and} \bibinfo{person}{Carol Fung}.} \bibinfo{year}{2024}\natexlab{c}.
\newblock \showarticletitle{{QUICwand: A Machine Learning Optimization-Based Hybrid Defense Approach Against QUIC Flooding Attacks}}. In \bibinfo{booktitle}{\emph{2024 20th International Conference on the Design of Reliable Communication Networks (DRCN)}}. \bibinfo{pages}{92--99}.
\newblock
\href{https://doi.org/10.1109/DRCN60692.2024.10539170}{doi:\nolinkurl{10.1109/DRCN60692.2024.10539170}}


\bibitem[Johnston(2023)]%
        {dpu-is-good}
\bibfield{author}{\bibinfo{person}{Rick Johnston}.} \bibinfo{year}{2023}\natexlab{}.
\newblock \bibinfo{title}{{Improving Data Center Energy Efficiency and Scalability}}.
\newblock
\urldef\tempurl%
\url{https://www.device42.com/blog/2023/12/14/improving-data-center-energy-efficiency-and-scalability}
\showURL{%
\tempurl}


\bibitem[Ke et~al\mbox{.}(2017)]%
        {lightgbm}
\bibfield{author}{\bibinfo{person}{Guolin Ke}, \bibinfo{person}{Qi Meng}, \bibinfo{person}{Thomas Finley}, \bibinfo{person}{Taifeng Wang}, \bibinfo{person}{Wei Chen}, \bibinfo{person}{Weidong Ma}, \bibinfo{person}{Qiwei Ye}, {and} \bibinfo{person}{Tie-Yan Liu}.} \bibinfo{year}{2017}\natexlab{}.
\newblock \showarticletitle{{LightGBM: a highly efficient gradient boosting decision tree}}. In \bibinfo{booktitle}{\emph{Proceedings of the 31st International Conference on Neural Information Processing Systems}} (Long Beach, California, USA) \emph{(\bibinfo{series}{NIPS'17})}. \bibinfo{publisher}{Curran Associates Inc.}, \bibinfo{address}{Red Hook, NY, USA}, \bibinfo{pages}{3149--3157}.
\newblock
\showISBNx{9781510860964}


\bibitem[Kim et~al\mbox{.}(2020)]%
        {dpu-tcp-tls}
\bibfield{author}{\bibinfo{person}{Duckwoo Kim}, \bibinfo{person}{SeungEon Lee}, {and} \bibinfo{person}{KyoungSoo Park}.} \bibinfo{year}{2020}\natexlab{}.
\newblock \showarticletitle{A Case for SmartNIC-accelerated Private Communication}. In \bibinfo{booktitle}{\emph{Proceedings of the 4th Asia-Pacific Workshop on Networking}} (Seoul, Republic of Korea) \emph{(\bibinfo{series}{APNet '20})}. \bibinfo{publisher}{Association for Computing Machinery}, \bibinfo{address}{New York, NY, USA}, \bibinfo{pages}{30--35}.
\newblock
\showISBNx{9781450388764}
\href{https://doi.org/10.1145/3411029.3411034}{doi:\nolinkurl{10.1145/3411029.3411034}}


\bibitem[Kim et~al\mbox{.}(2023a)]%
        {kim2023rearchitecting}
\bibfield{author}{\bibinfo{person}{Taehyun Kim}, \bibinfo{person}{Deondre~Martin Ng}, \bibinfo{person}{Junzhi Gong}, \bibinfo{person}{Youngjin Kwon}, \bibinfo{person}{Minlan Yu}, {and} \bibinfo{person}{KyoungSoo Park}.} \bibinfo{year}{2023}\natexlab{a}.
\newblock \showarticletitle{Rearchitecting the $\{$TCP$\}$ stack for $\{$I/O-Offloaded$\}$ content delivery}. In \bibinfo{booktitle}{\emph{20th USENIX Symposium on Networked Systems Design and Implementation (NSDI 23)}}. \bibinfo{pages}{275--292}.
\newblock


\bibitem[Kim et~al\mbox{.}(2023b)]%
        {iotcp}
\bibfield{author}{\bibinfo{person}{Taehyun Kim}, \bibinfo{person}{Deondre~Martin Ng}, \bibinfo{person}{Junzhi Gong}, \bibinfo{person}{Youngjin Kwon}, \bibinfo{person}{Minlan Yu}, {and} \bibinfo{person}{KyoungSoo Park}.} \bibinfo{year}{2023}\natexlab{b}.
\newblock \showarticletitle{Rearchitecting the {TCP} Stack for {I/O-Offloaded} Content Delivery}. In \bibinfo{booktitle}{\emph{20th USENIX Symposium on Networked Systems Design and Implementation (NSDI 23)}}. \bibinfo{publisher}{USENIX Association}, \bibinfo{address}{Boston, MA}, \bibinfo{pages}{275--292}.
\newblock
\showISBNx{978-1-939133-33-5}
\urldef\tempurl%
\url{https://www.usenix.org/conference/nsdi23/presentation/kim-taehyun}
\showURL{%
\tempurl}


\bibitem[Krawczyk et~al\mbox{.}(1997)]%
        {hmac}
\bibfield{author}{\bibinfo{person}{Hugo Krawczyk}, \bibinfo{person}{Mihir Bellare}, {and} \bibinfo{person}{Ran Canetti}.} \bibinfo{year}{1997}\natexlab{}.
\newblock \bibinfo{title}{{HMAC: Keyed-Hashing for Message Authentication}}.
\newblock \bibinfo{howpublished}{RFC 2104}.
\newblock
\href{https://doi.org/10.17487/RFC2104}{doi:\nolinkurl{10.17487/RFC2104}}


\bibitem[Krawczyk and Eronen(2010)]%
        {hkdf}
\bibfield{author}{\bibinfo{person}{Hugo Krawczyk} {and} \bibinfo{person}{Pasi Eronen}.} \bibinfo{year}{2010}\natexlab{}.
\newblock \bibinfo{title}{{HMAC-based Extract-and-Expand Key Derivation Function (HKDF)}}.
\newblock \bibinfo{howpublished}{RFC 5869}.
\newblock
\href{https://doi.org/10.17487/RFC5869}{doi:\nolinkurl{10.17487/RFC5869}}


\bibitem[Langley et~al\mbox{.}(2017)]%
        {gquic}
\bibfield{author}{\bibinfo{person}{Adam Langley}, \bibinfo{person}{Alistair Riddoch}, \bibinfo{person}{Alyssa Wilk}, \bibinfo{person}{Antonio Vicente}, \bibinfo{person}{Charles Krasic}, \bibinfo{person}{Dan Zhang}, \bibinfo{person}{Fan Yang}, \bibinfo{person}{Fedor Kouranov}, \bibinfo{person}{Ian Swett}, \bibinfo{person}{Janardhan Iyengar}, \bibinfo{person}{Jeff Bailey}, \bibinfo{person}{Jeremy Dorfman}, \bibinfo{person}{Jim Roskind}, \bibinfo{person}{Joanna Kulik}, \bibinfo{person}{Patrik Westin}, \bibinfo{person}{Raman Tenneti}, \bibinfo{person}{Robbie Shade}, \bibinfo{person}{Ryan Hamilton}, \bibinfo{person}{Victor Vasiliev}, \bibinfo{person}{Wan-Teh Chang}, {and} \bibinfo{person}{Zhongyi Shi}.} \bibinfo{year}{2017}\natexlab{}.
\newblock \showarticletitle{{The QUIC Transport Protocol: Design and Internet-Scale Deployment}}. In \bibinfo{booktitle}{\emph{Proceedings of the Conference of the ACM Special Interest Group on Data Communication}} \emph{(\bibinfo{series}{SIGCOMM '17})}. \bibinfo{publisher}{Association for Computing Machinery}, \bibinfo{address}{New York, NY, USA}, \bibinfo{pages}{183--196}.
\newblock


\bibitem[Li et~al\mbox{.}(2024a)]%
        {li2024triton}
\bibfield{author}{\bibinfo{person}{Xing Li}, \bibinfo{person}{Xiaochong Jiang}, \bibinfo{person}{Ye Yang}, \bibinfo{person}{Lilong Chen}, \bibinfo{person}{Yi Wang}, \bibinfo{person}{Chao Wang}, \bibinfo{person}{Chao Xu}, \bibinfo{person}{Yilong Lv}, \bibinfo{person}{Bowen Yang}, \bibinfo{person}{Taotao Wu}, {et~al\mbox{.}}} \bibinfo{year}{2024}\natexlab{a}.
\newblock \showarticletitle{Triton: A flexible hardware offloading architecture for accelerating apsara vswitch in alibaba cloud}. In \bibinfo{booktitle}{\emph{Proceedings of the ACM SIGCOMM 2024 Conference}}. \bibinfo{pages}{750--763}.
\newblock


\bibitem[Li et~al\mbox{.}(2024b)]%
        {bf3-acc-compress}
\bibfield{author}{\bibinfo{person}{Yuke Li}, \bibinfo{person}{Arjun Kashyap}, \bibinfo{person}{Weicong Chen}, \bibinfo{person}{Yanfei Guo}, {and} \bibinfo{person}{Xiaoyi Lu}.} \bibinfo{year}{2024}\natexlab{b}.
\newblock \showarticletitle{{ Accelerating Lossy and Lossless Compression on Emerging BlueField DPU Architectures }}. In \bibinfo{booktitle}{\emph{2024 IEEE International Parallel and Distributed Processing Symposium (IPDPS)}}. \bibinfo{publisher}{IEEE Computer Society}, \bibinfo{address}{Los Alamitos, CA, USA}, \bibinfo{pages}{373--385}.
\newblock


\bibitem[Liu et~al\mbox{.}(2021)]%
        {jaqen}
\bibfield{author}{\bibinfo{person}{Zaoxing Liu}, \bibinfo{person}{Hun Namkung}, \bibinfo{person}{Georgios Nikolaidis}, \bibinfo{person}{Jeongkeun Lee}, \bibinfo{person}{Changhoon Kim}, \bibinfo{person}{Xin Jin}, \bibinfo{person}{Vladimir Braverman}, \bibinfo{person}{Minlan Yu}, {and} \bibinfo{person}{Vyas Sekar}.} \bibinfo{year}{2021}\natexlab{}.
\newblock \showarticletitle{Jaqen: A {High-Performance} {Switch-Native} Approach for Detecting and Mitigating Volumetric {DDoS} Attacks with Programmable Switches}. In \bibinfo{booktitle}{\emph{30th USENIX Security Symposium (USENIX Security 21)}}. \bibinfo{publisher}{USENIX Association}, \bibinfo{pages}{3829--3846}.
\newblock
\showISBNx{978-1-939133-24-3}
\urldef\tempurl%
\url{https://www.usenix.org/conference/usenixsecurity21/presentation/liu-zaoxing}
\showURL{%
\tempurl}


\bibitem[Lychev et~al\mbox{.}(2015)]%
        {quic-analysis}
\bibfield{author}{\bibinfo{person}{Robert Lychev}, \bibinfo{person}{Samuel Jero}, \bibinfo{person}{Alexandra Boldyreva}, {and} \bibinfo{person}{Cristina Nita-Rotaru}.} \bibinfo{year}{2015}\natexlab{}.
\newblock \showarticletitle{{How Secure and Quick is QUIC? Provable Security and Performance Analyses}}. In \bibinfo{booktitle}{\emph{2015 IEEE Symposium on Security and Privacy}}. \bibinfo{pages}{214--231}.
\newblock
\href{https://doi.org/10.1109/SP.2015.21}{doi:\nolinkurl{10.1109/SP.2015.21}}


\bibitem[Mahan(2024)]%
        {dpu-is-good-3}
\bibfield{author}{\bibinfo{person}{Josh Mahan}.} \bibinfo{year}{2024}\natexlab{}.
\newblock \bibinfo{title}{{Data Processing Unit (DPU): Meaning and Role in Modern Computing}}.
\newblock \bibinfo{howpublished}{https://cc-techgroup.com/dpu/}.
\newblock


\bibitem[McGrew(2008)]%
        {aead}
\bibfield{author}{\bibinfo{person}{David McGrew}.} \bibinfo{year}{2008}\natexlab{}.
\newblock \bibinfo{title}{{An Interface and Algorithms for Authenticated Encryption}}.
\newblock \bibinfo{howpublished}{RFC 5116}.
\newblock
\href{https://doi.org/10.17487/RFC5116}{doi:\nolinkurl{10.17487/RFC5116}}


\bibitem[Nan et~al\mbox{.}(2025)]%
        {pno}
\bibfield{author}{\bibinfo{person}{Hailong Nan}, \bibinfo{person}{Zhe Zhou}, {and} \bibinfo{person}{Min Yang}.} \bibinfo{year}{2025}\natexlab{}.
\newblock \bibinfo{title}{Plug \& Offload: Transparently Offloading TCP Stack onto Off-path SmartNIC with PnO-TCP}.
\newblock
\showeprint[arxiv]{2503.22930}~[cs.DC]
\urldef\tempurl%
\url{https://arxiv.org/abs/2503.22930}
\showURL{%
\tempurl}


\bibitem[Nawrocki et~al\mbox{.}(2021)]%
        {quicsand}
\bibfield{author}{\bibinfo{person}{Marcin Nawrocki}, \bibinfo{person}{Raphael Hiesgen}, \bibinfo{person}{Thomas~C. Schmidt}, {and} \bibinfo{person}{Matthias W\"{a}hlisch}.} \bibinfo{year}{2021}\natexlab{}.
\newblock \showarticletitle{{QUICsand: quantifying QUIC reconnaissance scans and DoS flooding events}}. In \bibinfo{booktitle}{\emph{Proceedings of the 21st ACM Internet Measurement Conference}} \emph{(\bibinfo{series}{IMC '21})}. \bibinfo{publisher}{Association for Computing Machinery}, \bibinfo{address}{New York, NY, USA}, \bibinfo{pages}{283--291}.
\newblock


\bibitem[Nayak and Patgiri(2021)]%
        {cbf}
\bibfield{author}{\bibinfo{person}{Sabuzima Nayak} {and} \bibinfo{person}{Ripon Patgiri}.} \bibinfo{year}{2021}\natexlab{}.
\newblock \showarticletitle{{countBF: A General-purpose High Accuracy and Space Efficient Counting Bloom Filter}}. In \bibinfo{booktitle}{\emph{2021 17th International Conference on Network and Service Management (CNSM)}}. \bibinfo{pages}{355--359}.
\newblock


\bibitem[{NIST National Vulnerability Database} and {GitHub, Inc.}(2024)]%
        {cve-2024-53259}
\bibfield{author}{\bibinfo{person}{{NIST National Vulnerability Database}} {and} \bibinfo{person}{{GitHub, Inc.}}} \bibinfo{year}{2024}\natexlab{}.
\newblock \bibinfo{title}{{CVE-2024-53259}}.
\newblock \bibinfo{howpublished}{https://nvd.nist.gov/vuln/detail/CVE-2024-53259}.
\newblock


\bibitem[{NIST National Vulnerability Database} and {GitHub, Inc.}(2025a)]%
        {cve-2025-29785}
\bibfield{author}{\bibinfo{person}{{NIST National Vulnerability Database}} {and} \bibinfo{person}{{GitHub, Inc.}}} \bibinfo{year}{2025}\natexlab{a}.
\newblock \bibinfo{title}{{CVE-2025-29785}}.
\newblock \bibinfo{howpublished}{https://nvd.nist.gov/vuln/detail/CVE-2025-29785}.
\newblock


\bibitem[{NIST National Vulnerability Database} and {GitHub, Inc.}(2025b)]%
        {cve-2025-29908}
\bibfield{author}{\bibinfo{person}{{NIST National Vulnerability Database}} {and} \bibinfo{person}{{GitHub, Inc.}}} \bibinfo{year}{2025}\natexlab{b}.
\newblock \bibinfo{title}{{CVE-2025-29908}}.
\newblock \bibinfo{howpublished}{https://nvd.nist.gov/vuln/detail/CVE-2025-29908}.
\newblock


\bibitem[Novais and Verdi(2024)]%
        {dpu-ktls}
\bibfield{author}{\bibinfo{person}{Felipe A.~S. Novais} {and} \bibinfo{person}{Fábio~L. Verdi}.} \bibinfo{year}{2024}\natexlab{}.
\newblock \showarticletitle{Unlocking Security to the Board: An Evaluation of SmartNIC-driven TLS Acceleration with kTLS}. In \bibinfo{booktitle}{\emph{NOMS 2024-2024 IEEE Network Operations and Management Symposium}}. \bibinfo{pages}{1--9}.
\newblock
\href{https://doi.org/10.1109/NOMS59830.2024.10575053}{doi:\nolinkurl{10.1109/NOMS59830.2024.10575053}}


\bibitem[{NVIDIA Inc}(2024a)]%
        {doca}
\bibfield{author}{\bibinfo{person}{{NVIDIA Inc}}.} \bibinfo{year}{2024}\natexlab{a}.
\newblock \bibinfo{title}{{DOCA Documentation v2.7.0}}.
\newblock \bibinfo{howpublished}{\url{https://docs.nvidia.com/doca/archive/2-7-0/index.html}}.
\newblock


\bibitem[{NVIDIA Inc}(2024b)]%
        {bf3}
\bibfield{author}{\bibinfo{person}{{NVIDIA Inc}}.} \bibinfo{year}{2024}\natexlab{b}.
\newblock \bibinfo{title}{{NVIDIA BLUEFIELD-3 DPU}}.
\newblock \bibinfo{howpublished}{\url{https://www.nvidia.com/content/dam/en-zz/Solutions/Data-Center/documents/datasheet-nvidia-bluefield-3-dpu.pdf}}.
\newblock


\bibitem[{NVIDIA Inc}(2025)]%
        {dpa}
\bibfield{author}{\bibinfo{person}{{NVIDIA Inc}}.} \bibinfo{year}{2025}\natexlab{}.
\newblock \bibinfo{title}{{DPA Subsystem}}.
\newblock \bibinfo{howpublished}{\url{https://docs.nvidia.com/doca/sdk/dpa+subsystem/index.html}}.
\newblock


\bibitem[{quic-go orgnization}(2025)]%
        {quic-go}
\bibfield{author}{\bibinfo{person}{{quic-go orgnization}}.} \bibinfo{year}{2025}\natexlab{}.
\newblock \bibinfo{title}{{quic-go}}.
\newblock \bibinfo{howpublished}{https://github.com/quic-go/quic-go}.
\newblock


\bibitem[{QUIC Working Group}(2025)]%
        {quic-impl-list}
\bibfield{author}{\bibinfo{person}{{QUIC Working Group}}.} \bibinfo{year}{2025}\natexlab{}.
\newblock \bibinfo{title}{{QUIC Working Group - Implementations and tools}}.
\newblock \bibinfo{howpublished}{https://github.com/quicwg/quicwg.github.io/blob/main/implementations.md}.
\newblock


\bibitem[Rescorla(2018)]%
        {rfc8446}
\bibfield{author}{\bibinfo{person}{Eric Rescorla}.} \bibinfo{year}{2018}\natexlab{}.
\newblock \bibinfo{title}{{The Transport Layer Security (TLS) Protocol Version 1.3}}.
\newblock \bibinfo{howpublished}{RFC 8446}.
\newblock
\href{https://doi.org/10.17487/RFC8446}{doi:\nolinkurl{10.17487/RFC8446}}


\bibitem[Rescorla and Dierks(2008)]%
        {rfc5246}
\bibfield{author}{\bibinfo{person}{Eric Rescorla} {and} \bibinfo{person}{Tim Dierks}.} \bibinfo{year}{2008}\natexlab{}.
\newblock \bibinfo{title}{{The Transport Layer Security (TLS) Protocol Version 1.2}}.
\newblock \bibinfo{howpublished}{RFC 5246}.
\newblock
\href{https://doi.org/10.17487/RFC5246}{doi:\nolinkurl{10.17487/RFC5246}}


\bibitem[Robinson(2021)]%
        {dpu-is-good-2}
\bibfield{author}{\bibinfo{person}{Daniel Robinson}.} \bibinfo{year}{2021}\natexlab{}.
\newblock \bibinfo{title}{{Data processing units accelerate infrastructure performance}}.
\newblock
\urldef\tempurl%
\url{https://www.techtarget.com/searchdatacenter/feature/Data-processing-units-accelerate-infrastructure-performance}
\showURL{%
\tempurl}


\bibitem[Salowey et~al\mbox{.}(2008)]%
        {aes-gcm}
\bibfield{author}{\bibinfo{person}{Joseph~A. Salowey}, \bibinfo{person}{David McGrew}, {and} \bibinfo{person}{Abhijit Choudhury}.} \bibinfo{year}{2008}\natexlab{}.
\newblock \bibinfo{title}{{AES Galois Counter Mode (GCM) Cipher Suites for TLS}}.
\newblock \bibinfo{howpublished}{RFC 5288}.
\newblock
\href{https://doi.org/10.17487/RFC5288}{doi:\nolinkurl{10.17487/RFC5288}}


\bibitem[Sengupta et~al\mbox{.}(2025)]%
        {quic-attack-survey}
\bibfield{author}{\bibinfo{person}{Jayasree Sengupta}, \bibinfo{person}{Debasmita Dey}, \bibinfo{person}{Simone Ferlin-Reiter}, \bibinfo{person}{Nirnay Ghosh}, {and} \bibinfo{person}{Vaibhav Bajpai}.} \bibinfo{year}{2025}\natexlab{}.
\newblock \bibinfo{title}{{How Resilient is QUIC to Security and Privacy Attacks?}}
\newblock
\showeprint[arxiv]{2401.06657}~[cs.CR]
\urldef\tempurl%
\url{https://arxiv.org/abs/2401.06657}
\showURL{%
\tempurl}


\bibitem[Teyssier et~al\mbox{.}(2023)]%
        {quicshield}
\bibfield{author}{\bibinfo{person}{Benjamin Teyssier}, \bibinfo{person}{Y~A Joarder}, {and} \bibinfo{person}{Carol Fung}.} \bibinfo{year}{2023}\natexlab{}.
\newblock \showarticletitle{{QUICShield: A Rapid Detection Mechanism Against QUIC-Flooding Attacks}}. In \bibinfo{booktitle}{\emph{2023 IEEE Virtual Conference on Communications (VCC)}}. \bibinfo{pages}{43--48}.
\newblock
\href{https://doi.org/10.1109/VCC60689.2023.10474735}{doi:\nolinkurl{10.1109/VCC60689.2023.10474735}}


\bibitem[Thomson and Turner(2021)]%
        {rfc9001}
\bibfield{author}{\bibinfo{person}{Martin Thomson} {and} \bibinfo{person}{Sean Turner}.} \bibinfo{year}{2021}\natexlab{}.
\newblock \bibinfo{title}{{Using TLS to Secure QUIC}}.
\newblock \bibinfo{howpublished}{RFC 9001}.
\newblock
\href{https://doi.org/10.17487/RFC9001}{doi:\nolinkurl{10.17487/RFC9001}}


\bibitem[{TurboRetry}(2025)]%
        {turboretry-repo}
\bibfield{author}{\bibinfo{person}{{TurboRetry}}.} \bibinfo{year}{2025}\natexlab{}.
\newblock \bibinfo{title}{{TurboRetry}}.
\newblock
\urldef\tempurl%
\url{https://github.com/wujiahao15/TurboRetry}
\showURL{%
\tempurl}


\bibitem[Wu et~al\mbox{.}(2019)]%
        {bayesian-opt1}
\bibfield{author}{\bibinfo{person}{Jia Wu}, \bibinfo{person}{Xiu-Yun Chen}, \bibinfo{person}{Hao Zhang}, \bibinfo{person}{Li-Dong Xiong}, \bibinfo{person}{Hang Lei}, {and} \bibinfo{person}{Si-Hao Deng}.} \bibinfo{year}{2019}\natexlab{}.
\newblock \showarticletitle{{Hyperparameter Optimization for Machine Learning Models Based on Bayesian Optimizationb}}.
\newblock \bibinfo{journal}{\emph{Journal of Electronic Science and Technology}} \bibinfo{volume}{17}, \bibinfo{number}{1} (\bibinfo{year}{2019}), \bibinfo{pages}{26--40}.
\newblock
\showISSN{1674-862X}
\href{https://doi.org/10.11989/JEST.1674-862X.80904120}{doi:\nolinkurl{10.11989/JEST.1674-862X.80904120}}


\bibitem[Wu et~al\mbox{.}(2025)]%
        {lemon}
\bibfield{author}{\bibinfo{person}{Wenhao Wu}, \bibinfo{person}{Zhenyu Li}, \bibinfo{person}{Xilai Liu}, \bibinfo{person}{Zhaohua Wang}, \bibinfo{person}{Heng Pan}, \bibinfo{person}{Guangxing Zhang}, {and} \bibinfo{person}{Gaogang Xie}.} \bibinfo{year}{2025}\natexlab{}.
\newblock \showarticletitle{Lemon: Network-Wide DDoS Detection with Routing-Oblivious Per-Flow Measurement}. In \bibinfo{booktitle}{\emph{34th USENIX Security Symposium (USENIX Security 25)}}. \bibinfo{publisher}{USENIX Association}.
\newblock


\bibitem[Xie et~al\mbox{.}(2021)]%
        {sqcusum}
\bibfield{author}{\bibinfo{person}{Liyan Xie}, \bibinfo{person}{Shaofeng Zou}, \bibinfo{person}{Yao Xie}, {and} \bibinfo{person}{Venugopal~V. Veeravalli}.} \bibinfo{year}{2021}\natexlab{}.
\newblock \showarticletitle{{Sequential (Quickest) Change Detection: Classical Results and New Directions}}.
\newblock \bibinfo{journal}{\emph{IEEE Journal on Selected Areas in Information Theory}} \bibinfo{volume}{2}, \bibinfo{number}{2} (\bibinfo{year}{2021}), \bibinfo{pages}{494--514}.
\newblock
\href{https://doi.org/10.1109/JSAIT.2021.3072962}{doi:\nolinkurl{10.1109/JSAIT.2021.3072962}}


\bibitem[Xing et~al\mbox{.}(2021)]%
        {ripple}
\bibfield{author}{\bibinfo{person}{Jiarong Xing}, \bibinfo{person}{Wenqing Wu}, {and} \bibinfo{person}{Ang Chen}.} \bibinfo{year}{2021}\natexlab{}.
\newblock \showarticletitle{Ripple: A Programmable, Decentralized {Link-Flooding} Defense Against Adaptive Adversaries}. In \bibinfo{booktitle}{\emph{30th USENIX Security Symposium (USENIX Security 21)}}. \bibinfo{publisher}{USENIX Association}, \bibinfo{pages}{3865--3881}.
\newblock
\shownote{https://www.usenix.org/conference/usenixsecurity21/presentation/xing}.
\newblock
\showISBNx{978-1-939133-24-3}
\urldef\tempurl%
\url{https://www.usenix.org/conference/usenixsecurity21/presentation/xing}
\showURL{%
\tempurl}


\bibitem[Yang et~al\mbox{.}(2020)]%
        {quic-nic-offload}
\bibfield{author}{\bibinfo{person}{Xiangrui Yang}, \bibinfo{person}{Lars Eggert}, \bibinfo{person}{J\"{o}rg Ott}, \bibinfo{person}{Steve Uhlig}, \bibinfo{person}{Zhigang Sun}, {and} \bibinfo{person}{Gianni Antichi}.} \bibinfo{year}{2020}\natexlab{}.
\newblock \showarticletitle{{Making QUIC Quicker With NIC Offload}}. In \bibinfo{booktitle}{\emph{Proceedings of the Workshop on the Evolution, Performance, and Interoperability of QUIC}} (Virtual Event, USA) \emph{(\bibinfo{series}{EPIQ '20})}. \bibinfo{publisher}{Association for Computing Machinery}, \bibinfo{address}{New York, NY, USA}, \bibinfo{pages}{21--27}.
\newblock
\showISBNx{9781450380478}
\href{https://doi.org/10.1145/3405796.3405827}{doi:\nolinkurl{10.1145/3405796.3405827}}


\bibitem[Yoo et~al\mbox{.}(2024)]%
        {smartcookie}
\bibfield{author}{\bibinfo{person}{Sophia Yoo}, \bibinfo{person}{Xiaoqi Chen}, {and} \bibinfo{person}{Jennifer Rexford}.} \bibinfo{year}{2024}\natexlab{}.
\newblock \showarticletitle{{SmartCookie}: Blocking {Large-Scale} {SYN} Floods with a {Split-Proxy} Defense on Programmable Data Planes}. In \bibinfo{booktitle}{\emph{33rd USENIX Security Symposium (USENIX Security 24)}}. \bibinfo{publisher}{USENIX Association}, \bibinfo{address}{Philadelphia, PA}, \bibinfo{pages}{217--234}.
\newblock
\showISBNx{978-1-939133-44-1}
\urldef\tempurl%
\url{https://www.usenix.org/conference/usenixsecurity24/presentation/yoo}
\showURL{%
\tempurl}


\bibitem[Zhang et~al\mbox{.}(2020)]%
        {poseidon}
\bibfield{author}{\bibinfo{person}{Menghao Zhang}, \bibinfo{person}{Guanyu Li}, \bibinfo{person}{Shicheng Wang}, \bibinfo{person}{Chang Liu}, \bibinfo{person}{Ang Chen}, \bibinfo{person}{Hongxin Hu}, \bibinfo{person}{Guofei Gu}, \bibinfo{person}{Qi Li}, \bibinfo{person}{Mingwei Xu}, {and} \bibinfo{person}{Jianping Wu}.} \bibinfo{year}{2020}\natexlab{}.
\newblock \showarticletitle{Poseidon: Mitigating Volumetric DDoS Attacks with Programmable Switches}. In \bibinfo{booktitle}{\emph{27th Annual Network and Distributed System Security Symposium, {NDSS} 2020, San Diego, California, USA, February 23-26, 2020}}. \bibinfo{publisher}{The Internet Society}.
\newblock
\href{https://doi.org/10.14722/ndss.2020.24007}{doi:\nolinkurl{10.14722/ndss.2020.24007}}


\bibitem[Zhao et~al\mbox{.}(2020)]%
        {pigasus}
\bibfield{author}{\bibinfo{person}{Zhipeng Zhao}, \bibinfo{person}{Hugo Sadok}, \bibinfo{person}{Nirav Atre}, \bibinfo{person}{James~C. Hoe}, \bibinfo{person}{Vyas Sekar}, {and} \bibinfo{person}{Justine Sherry}.} \bibinfo{year}{2020}\natexlab{}.
\newblock \showarticletitle{Achieving 100Gbps Intrusion Prevention on a Single Server}. In \bibinfo{booktitle}{\emph{14th USENIX Symposium on Operating Systems Design and Implementation (OSDI 20)}}. \bibinfo{publisher}{USENIX Association}, \bibinfo{pages}{1083--1100}.
\newblock
\showISBNx{978-1-939133-19-9}


\end{thebibliography}

\appendix

\section{Open Science}
We adhere to the open science policy and make the prototype of \sysname publicly available at~\cite{turboretry-repo}.
The repository includes the following key components:
\begin{itemize}
    \item \textbf{\sysname Source Code:} The complete implementation of the proposed DPU-based offloading prototype.
    \item \textbf{Host-side Integration:} Patches for standard QUIC implementations (\texttt{aioquic} and \texttt{quiche}) enabling the cooperative \retry mechanism.
    \item \textbf{Attack Tool:} A high-performance traffic generator capable of synthesizing large-scale QUIC handshake floods.
    \item \textbf{Client Tool:} Scripts and tools for generating legitimate background HTTP/3 traffic.
    \item \textbf{Documentation:} Guides written in Markdown, namely \texttt{README.md} files, for compiling and running the artifact.
\end{itemize}
\section{Ethical Considerations}
Our study focuses on accelerating and securing the QUIC Retry mechanism by offloading critical operations to a data processing unit. 
We conduct a stakeholder-based ethics analysis to identify potential benefits and risks. The key stakeholders include:
\begin{itemize}
    \item \textbf{End-users} of QUIC-based services. They can benefit from stronger protection against handshake flooding attacks and experience more robust service.
    \item \textbf{Cloud service providers and network operators}. They gain improved scalability and security efficiency for QUIC-based services, but must carefully manage deployment complexity and hardware costs.
    \item \textbf{The broader research and security community}. They may benefit from the split design insights and open discussion of offloading techniques, but also face the risk of adversaries adapting their attack strategies based on disclosed details.
    \item \textbf{Standards organizations.} They may incorporate insights from our work into protocol evolution, but must balance interoperability with deployment feasibility.
\end{itemize}

To mitigate potential negative outcomes, we ensure that all experiments were conducted in our controlled testbeds without real-world damage. 
Overall, our research does not raise additional ethical concerns beyond the controlled testbed setting.

\section{Token Formats and Lifetimes of Existing QUIC implementations}

\newtext{
Table~\ref{apptab:quic-token-summary} summarizes our survey of Retry-token handling in the open-source QUIC implementations listed by the QUIC working group~\cite{quic-impl-list}. 
We examine the token format, protection mechanism, and lifetime used by each implementation. 
The results show that Retry-token handling is implementation-specific: different implementations encode different fields and adopt different expiration policies. 
At the same time, most implementations protect tokens cryptographically and use them to carry address-validation state. 
}

\begin{table*}[t]
    \caption{\newtext{Token encryption, formats, binding methods, and lifetimes in existing QUIC implementations}}
    \label{apptab:quic-token-summary}
    \centering
    \footnotesize
    \setlength{\tabcolsep}{2.5pt}
    \renewcommand{\arraystretch}{1.15}
    \begin{tabular}{
        p{0.1\textwidth}
        p{0.105\textwidth}
        p{0.345\textwidth}
        p{0.17\textwidth}
        p{0.205\textwidth}
    }
        \toprule
        \textbf{Implementation} &
        \textbf{\makecell[l]{Encryption\\Algorithm}} &
        \textbf{Token format and major fields} &
        \textbf{AAD binding information (AES-GCM specific)} &
        \textbf{Lifetime and implementation characteristics} \\
        \midrule

        aioquic &
        RSA &
        \texttt{Encrypted(Client IP || ODCID || RSCID)} &
        Implicitly protected by the RSA-encrypted payload, without AAD. &
        Server-maintained RSA key; asymmetric protection of token payload. \\

        Haskwell quic &
        AES-GCM-256 &
        \texttt{Masked Header(10B) || Encrypted(content) || Auth Tag}.
        The content includes
        \texttt{QUIC Version || Token\_LifeTime || Timestamp || DCID || SCID || ODCID}. &
        Empty AAD. &
        Token includes QUIC version, token lifetime, timestamp, and connection IDs. \\

        Linuxquic &
        AES-GCM-128 &
        \texttt{AAD(Flag || Client IP) || Encrypted(Timestamp || ODCID) || Auth Tag}.
        The flag distinguishes regular tokens from retry tokens. &
        \texttt{Flag || Client IP} &
        Retry token: 3 s; NEW\_TOKEN: 10 s; binding to token usage and client IP. \\

        HAProxy &
        AES-GCM-128 &
        \texttt{Flag(1B) || Encrypted(ODCIDL || ODCID || Timestamp(4B)) || Auth Tag || Salt}.
        Salt is a plaintext field used with the server key to derive the AES key and nonce. &
        \texttt{QUIC Version || Client IP || Client Port || DCID} &
        Retry token: 10 s; NEW\_TOKEN: 7 days; binding to version, client address, port, and DCID. \\

        lsquic &
        AES-GCM-128 &
        \texttt{Nonce(12B) || Encrypted(content) || Auth Tag(16B)}.
        The content includes
        \texttt{TokenVersion || Timestamp || IPPROTO || Client IP || Client Port || ODCID}.
        IPPROTO distinguishes IPv4 from IPv6. &
        \texttt{DCID} &
        Retry token: 600 s; binding to DCID. \\

        MsQuic &
        AES-GCM-128/256 &
        \texttt{AAD(Flag(1b) || Timestamp(63b)) || Encrypted(Client IP || ODCIDL || ODCID) || Auth Tag(16B)}.
        The flag indicates whether the token is a NEW\_TOKEN. &
        \texttt{Flag || Timestamp} &
        Key rotation: 30 s; token lifetime: 60 s; authenticated timestamp in AAD. \\

        mvfst &
        AES-GCM-128 &
        \texttt{Nonce(12B) || Encrypted(Timestamp)(8B) || Auth Tag(16B)} &
        \texttt{``RetryToken" || ODCID || Client IP} &
        Retry token: 5 min; AAD-level binding to original DCID and client IP. \\

        quic-go &
        AES-GCM-256 &
        \texttt{Nonce(32B) || Encrypted(ASN.1-encoded content) || Auth Tag(16B)}.
        The encoded content includes
        \texttt{IsRetryToken || Client IP || Timestamp || ODCID || RSCID}. &
        Empty AAD. &
        No AAD binding; prefix-based client-IP encoding for IP/string distinction. \\

        quinn &
        AES-GCM-256 &
        \texttt{Encrypted(content) || Auth Tag(16B) || Nonce(16B)}.
        The content includes
        \texttt{Type(1B) || IPPROTO || Client IP || Client Port || ODCIDL || ODCID || Timestamp(8B)}.
        Type \texttt{0x00} denotes Retry; IPPROTO values 0 and 1 denote IPv4 and IPv6, respectively. &
        Empty AAD. &
        Retry token: 15 s; NEW\_TOKEN: 14 days; binding fields carried in encrypted token content. \\

        neqo &
        AES-GCM-128 &
        \texttt{Header || Encrypted(content) || Auth Tag}.
        The header includes
        \texttt{Identifier(5B) || Version(1B) || KeyID || Salt(16B)}.
        Identifier distinguishes Retry from NEW\_TOKEN.
        The encrypted content includes \texttt{Expiration || ODCID}. &
        \texttt{Identifier || IPPROTO || Client IP || Client Port} &
        Retry token: 5 s; NEW\_TOKEN: 24 h; binding to token type, client address, and client port. \\

        ngtcp2 &
        AES-GCM-128 &
        \texttt{Magic(1B) || Encrypted(content) || Auth Tag(16B) || Salt(16B)}.
        Magic is \texttt{0xB7}.
        The content is \texttt{Client IP || ODCIDL || ODCID || Timestamp(8B)}. &
        \texttt{Version(4B) || RSCID} &
        Binding to QUIC version and RSCID; fixed lifetime not exposed in this table. \\

        picoquic &
        AES-GCM-128 &
        \texttt{RandomSequence(8B) || Encrypted(content) || Auth Tag}.
        The most significant bit of RandomSequence is a flag, and the remaining 63 bits are random.
        The flag has the same meaning as in neqo.
        The content includes
        \texttt{Expire Time(8B) || ODCID || RSCID || Initial PN || Zero Padding}. &
        \texttt{Client IP} &
        Retry token: 2 min; NEW\_TOKEN: 24 h; binding to client IP. \\

        quicly &
        AES-GCM-128 &
        \texttt{Type(1B) || IV || Encrypted(content) || Auth Tag}.
        Type has the same meaning as the flag in picoquic.
        The content includes
        \texttt{Timestamp(8B) || Client IP || Client Port || ODCID || SCID || RSCID || App Data}.
        App Data is optional application-defined data. &
        Optional AAD, typically \texttt{Type || IV}. &
        Retry token: 30 s; NEW\_TOKEN: 10 min; optional AAD with token type and IV. \\

        xquic &
        AES-GCM-128 &
        \texttt{Random IV(12B) || Flag(1B) || Encrypted(Client IP || Expire Time) || Auth Tag(16B)}.
        The upper 6 bits of Flag indicate the IP type, where 0 denotes IPv4 and 1 denotes IPv6;
        the lower 2 bits indicate the key-rotation version. &
        No AAD. &
        Retry token: 5 s; NEW\_TOKEN: 7 days; no AAD binding; IP type and key version encoded in flag. \\

        Cloudflare Quiche &
        Application-defined; examples use no protection or HMAC-SHA256 &
        The \texttt{quiche-server.rs} example uses
        \texttt{"quiche"(6B) || Client IP || ODCID}, without encryption or cryptographic authentication.
        The \texttt{tokio-quiche} example uses
        \texttt{HMAC-SHA256 Tag(32B) || Client IP || Original DCID}. &
        Application-defined. \texttt{tokio-quiche} uses HMAC-SHA256 to authenticate the client address and ODCID. &
        Application-defined token handling; examples from unprotected tokens to HMAC-authenticated tokens. \\

        Google Quiche &
        Application-defined &
        The NEW\_TOKEN format is
        \texttt{Prefix(0x00) || Nonce(12B) || Encrypted(content) || Auth Tag},
        where the content includes \texttt{Client IP || Timestamp || Optional Parameters}. &
        Application-defined. &
        NEW\_TOKEN: 86400 s; application-side Retry token generation; AES-256-GCM protection for NEW\_TOKEN. \\

        kwik &
        None &
        \texttt{Random Token(37B)}.
        The server generates a 37-byte random token using \texttt{SecureRandom}. &
        No encryption, no AAD field. &
        No cryptographic binding; server-side token storage and comparison. \\

        \bottomrule
    \end{tabular}
\end{table*}

\end{document}